\documentclass[useAMS,usenatbib]{mn2e}

\usepackage{graphicx}
\usepackage{amsmath}
\usepackage{amssymb}
\usepackage{natbib}
\usepackage{aas_macros}
\usepackage{color}
\usepackage{float}
\usepackage{txfonts}

\usepackage{hyperref}
\hypersetup{
    colorlinks,%
    citecolor=blue,%
    linkcolor=blue,%
    urlcolor=blue
}

\title[Pulsar lightcurves from first principles]{Modeling high-energy pulsar lightcurves from first principles}

\author[Beno\^it Cerutti et al.]{Beno\^it Cerutti$^{1,2,3}$\thanks{E-mail: benoit.cerutti@obs.ujf-grenoble.fr}\thanks{Lyman Spitzer Jr. Fellow.}, Alexander A. Philippov$^{1}$ and Anatoly Spitkovsky$^{1}$\\
$^{1}$Department of Astrophysical Sciences, Princeton University, Princeton, NJ 08544, USA\\
$^{2}$Univ. Grenoble Alpes, IPAG, F-38000 Grenoble, France\\
$^{3}$CNRS, IPAG, F-38000 Grenoble, France}

\date{Accepted --. Received --; in original form --}

\pubyear{2016}

\begin{document}
\label{firstpage}
\pagerange{\pageref{firstpage}--\pageref{lastpage}}
\maketitle

\begin{abstract}
Current models of gamma-ray lightcurves in pulsars suffer from large uncertainties on the precise location of particle acceleration and radiation. Here, we present an attempt to alleviate these difficulties by solving for the electromagnetic structure of the oblique magnetosphere, particle acceleration, and the emission of radiation self-consistently, using 3D spherical particle-in-cell simulations. We find that the low-energy radiation is synchro-curvature radiation from the polar-cap regions within the light cylinder. In contrast, the high-energy emission is synchrotron radiation that originates exclusively from the Y-point and the equatorial current sheet where relativistic magnetic reconnection accelerates particles. In most cases, synthetic high-energy lightcurves contain two peaks that form when the current sheet sweeps across the observer's line of sight. We find clear evidence of caustics in the emission pattern from the current sheet. High-obliquity solutions can present up to two additional secondary peaks from energetic particles in the wind region accelerated by the reconnection-induced flow near the current sheet. The high-energy radiative efficiency depends sensitively on the viewing angle, and decreases with increasing pulsar inclination. The high-energy emission is concentrated in the equatorial regions where most of the pulsar spindown is released and dissipated. These results have important implications for the interpretation of gamma-ray pulsar data.
\end{abstract}


%%%%%%%%%%%%%%%%%%%%%%%%%%%%%%%%%%%%%%%%%%%%%%%%%%

%%%%%%%%%%%%%%%%% BODY OF PAPER %%%%%%%%%%%%%%%%%%

\section{Introduction}

The high-energy radiation from pulsars is characterized by short bright pulses modulated with the stellar rotation period \citep{2010ApJS..187..460A, 2013ApJS..208...17A}, which results most likely from the misalignment between the rotation axis and the magnetic axis of the star. Each known lightcurve is unique and constitutes a real fingerprint for each pulsar. Although all different, the majority of lightcurves present similar features, most notably the double-peaked structure, often with significant emission in between both peaks (the bridge emission). Extensive theoretical efforts have been concentrated on understanding the shape of pulsar gamma-ray lightcurves, with the ultimate hope that the structure of pulsar magnetospheres could be reverse-engineered from them.

It is commonly accepted that the gamma-ray emission originates somewhere between the neutron star surface and the light-cylinder radius where the co-rotating velocity equals the speed of light. However, there are still large uncertainties on the exact location of particle acceleration and radiation in the magnetosphere. In current magnetospheric models, particle acceleration occurs in small ad-hoc regions where the plasma density is low, such that a strong unscreened electric field can be present. The usual suspected locations for these gaps are: (i) near the star at the base of the open field lines (polar-cap model, \citealt{1971ApJ...164..529S, 1975ApJ...196...51R, 1978ApJ...225..226H, 1982ApJ...252..337D}) (ii) in the region between the null-surface (defined where the Goldreich-Julian charge density goes to zero, \citealt{1969ApJ...157..869G}) and the last open field lines (outer-gap model, \citealt{1986ApJ...300..500C, 1986ApJ...300..522C, 1995ApJ...438..314R}), and (iii) along the separatrix current layers at the boundary between the closed and open field lines, extending from the stellar surface up to the light-cylinder in some cases (slot-gap \citealt{1979ApJ...231..854A, 1983ApJ...266..215A, 2003ApJ...588..430M, 2004ApJ...606.1143M}, two-pole caustics \citealt{2003ApJ...598.1201D} models), but the {\em Fermi}-LAT data currently favor emission from the outer magnetosphere (e.g., \citealt{2010ApJS..187..460A}). Outer-gap models usually assume a pure dipolar geometry for the field lines. The deviations from the more realistic force-free configuration change the expected radiative signatures and reduce the predictive power of these models \citep{2010ApJ...715.1270B}. Working directly with the force-free fields gives new insight into the formation of gamma-ray lightcurves by tracing potential emitting field lines \citep{2010ApJ...715.1282B}, or using resistive force-free fields and test particles \citep{2012ApJ...754L...1K, 2014ApJ...793...97K}. These approaches provide a more self-consistent picture, but it still suffers from uncertainties on the location of accelerating zones since, by construction, there is no unscreened electric field in an ideal MHD force-free model. Hence, one has to assume where non-ideal effects should occur by prescribing an arbitrary resistivity in the magnetosphere \citep{2012ApJ...746...60L, 2012ApJ...754L...1K}, which is not fully satisfying.

The recent progress in global particle-in-cell (PIC) simulations of pulsar magnetospheres has brought a better understanding of plasma generation, the location of non-ideal regions and particle acceleration in pulsars \citep{2014ApJ...785L..33P, 2014ApJ...795L..22C, 2015MNRAS.448..606C, 2015ApJ...801L..19P, 2015MNRAS.449.2759B}. One of the main findings of these investigations is the key role of reconnection for particle acceleration within the equatorial current sheet that forms beyond the light cylinder in between the two magnetic polarities \citep{1990ApJ...349..538C}. These results suggest that the current sheet could be at the origin of the gamma-ray emission \citep{1996A&A...311..172L, 2002A&A...388L..29K, 2012MNRAS.424.2023P, 2013A&A...550A.101A, 2014ApJ...780....3U, 2015MNRAS.449L..51M}, but the radiative signature has not been clearly established from the PIC simulations. In this study, we report on an attempt to model the structure of the oblique magnetosphere, particle acceleration and, most importantly here, the emission of radiation all together and self-consistently, using global three-dimensional (3D) spherical PIC simulations. The radiation reaction force on the dynamics of particles is taken into account accordingly. This work focuses on the plasma-filled magnetosphere, which is most relevant to young gamma-ray pulsars.

Our main objectives are to (i) unambiguously identify the location of particle acceleration and radiation in the magnetosphere, (ii) characterize the nature of the emission, and (iii) deduce observables (lightcurves and spectra) self-consistently from the simulations, as function of the angle between the spin and the magnetic axis of the star (hereafter, the obliquity angle, $\chi$). Our goal is not to fit observations at this point, but instead, we provide a proof of principle that the PIC approach is suitable for solving this problem. In the following, we present the numerical techniques and procedures used in this study (Sect.~\ref{sect_setup}), with a particular emphasis on the treatment of the radiation reaction force and the emission of photons in the PIC simulations (Sect.~\ref{sect_rad}). The results are described in Sect.~\ref{sect_results} and discussed in Sect.~\ref{sect_conclusion}.

\section{Numerical setup}\label{sect_setup}

\begin{figure}
\centering
\includegraphics[width=8cm]{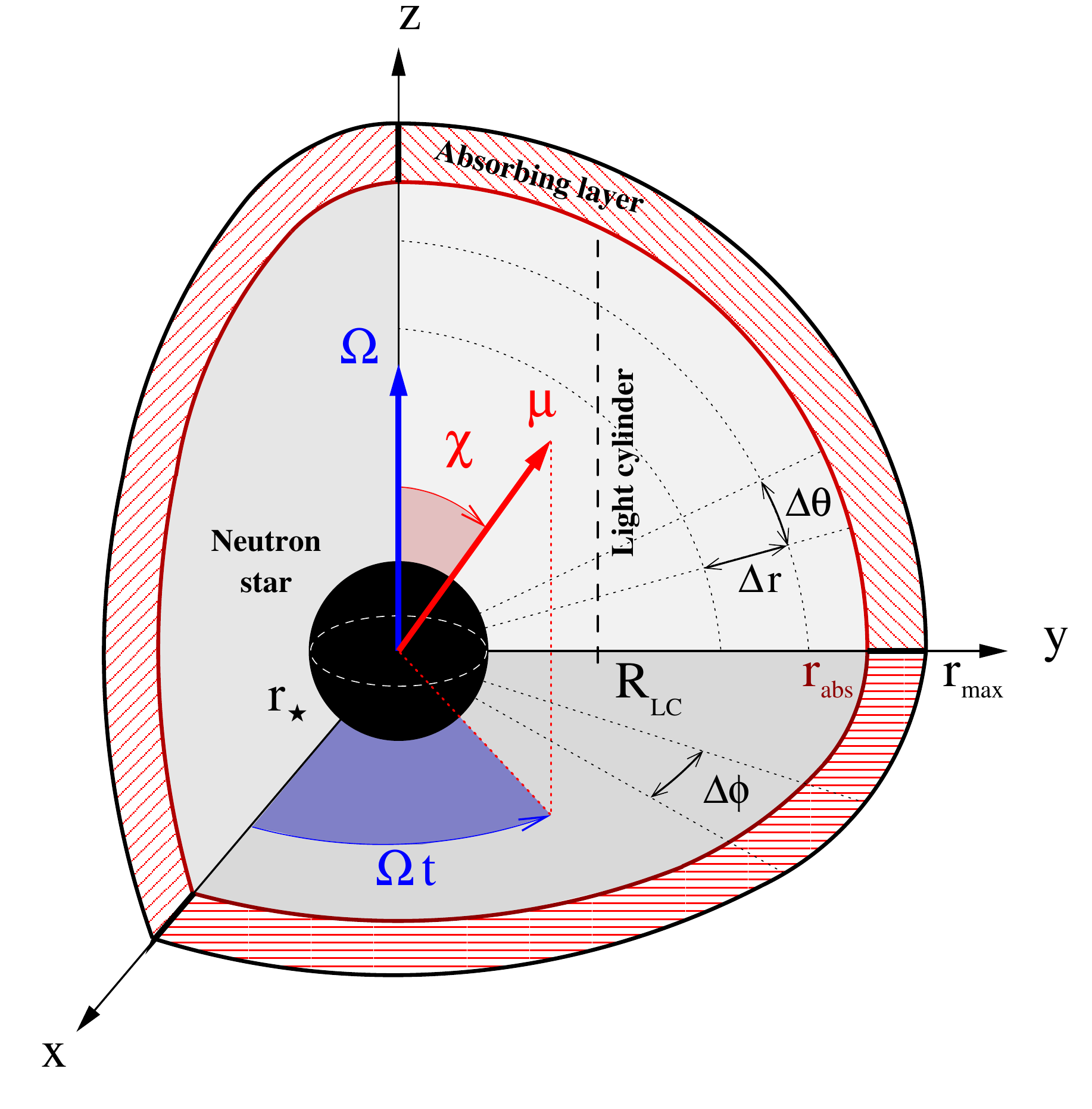}
\caption{Geometrical setup used in this study for the 3D PIC modeling of the misaligned pulsar magnetosphere. The angle between the magnetic moment of the star ($\boldsymbol{\mu}$) and the spin axis ($\boldsymbol{\Omega}$) is the obliquity angle $\chi$. The box is a spherical shell of inner radius $r_{\rm min}$ set at the neutron star radius, $r_{\star}$, and of outer radius $r_{\rm max}=10 r_{\star}$. The outer boundary is coated with a layer of thickness $\left(r_{\rm max}-r_{\rm abs}\right)=r_{\star}$ that absorbs all electromagnetic waves and particles leaving the domain. The grid cells are logarithmically spaced in $r$, and uniformly in $\theta$ and $\phi$.}
\label{fig_geo}
\end{figure}

We use the relativistic PIC code {\tt ZELTRON} \citep{2013ApJ...770..147C} which was recently upgraded to handle non-uniform 2D axisymmetric spherical grids in the context of aligned pulsars \citep{2015MNRAS.448..606C}. To model misaligned rotators, we have extended the spherical grid to full 3D, where the usual $(r,\theta,\phi)$ spherical coordinate system is used throughout this paper (Figure~\ref{fig_geo}). Maxwell's equations are solved on the 3D spherical Yee-mesh using their cell-integrated expressions as in \citet{2015MNRAS.448..606C} (for completeness, their general 3D forms are reported here in the Appendix). The particle motion is solved on a regular Cartesian grid using the modified Boris push by \citet{2010NJPh...12l3005T} to account for the radiation reaction force (see Sect.~\ref{sect_frad}). The particle positions and velocities are remapped every time step to the spherical grid for charge and current depositions using the volume weighting technique (a trilinear interpolation in $r^3$, $\cos\theta$ and $\phi$).

The computational domain is a spherical shell whose inner radius coincides with the neutron star surface, i.e., $r_{\rm min}=r_{\star}$, and extends up to $r_{\rm max}=10 r_{\star}$. The light-cylinder radius is set at $R_{\rm LC}=3 r_{\star}$. The shell covers the full $4\pi$ steradians, i.e., with $\theta\in\left[0,\pi\right]$ and $\phi\in\left[0,2\pi\right]$. The grid points are logarithmically spaced in radius and uniformly spaced in $\theta$ and $\phi$.

Each simulation is initialized with a rotating dipole in vacuum, whose magnetic moment $\boldsymbol{\mu}$ is inclined at an angle $\chi$ with respect to the rotation axis (where $\theta=0^{\rm o}$, see Figure~\ref{fig_geo}). The magnetic field components of a dipole spinning at the angular velocity $\Omega=c/R_{\rm LC}$, where $c$ is the speed of light, are given by
\begin{eqnarray}
\mathbf{B_{\rm r}} \hspace{-0.2cm}&=&\hspace{-0.2cm} \frac{2\mu}{r^3}\left[\sin\chi\sin\theta\cos\left(\Omega t-\phi\right)+\cos\chi\cos\theta\right]\mathbf{e_{r}}\hspace{1.5cm} \label{eq_br}\\
\mathbf{B_{\theta}} \hspace{-0.25cm}&=&\hspace{-0.2cm} \frac{\mu}{r^3}\left[-\sin\chi\cos\theta\cos\left(\Omega t-\phi\right)+\cos\chi\sin\theta\right]\mathbf{e_{\theta}}\\
\mathbf{B_{\phi}} \hspace{-0.3cm}&=&\hspace{-0.2cm} -\frac{\mu}{r^3}\sin\chi\sin\left(\Omega t-\phi\right)\mathbf{e_{\phi}}.
\label{eq_dipole}
\end{eqnarray}
At time $t=0$, the magnetic field is set by the above equations everywhere in space. As the simulation proceeds, we enforce the magnetic field to follow the rotating dipole only at the neutron star surface, i.e., at $r=r_{\star}$. On the Yee lattice (see Figure~\ref{fig_3dyee}), only $\mathbf{B_{\rm r}}$ is on the stellar surface and needs to be updated with Eq.~(\ref{eq_br}) every time step. The fast rotation of the magnetic field lines induces a poloidal electric field at $r=r_{\star}$ given by
\begin{equation}
\mathbf{E}=-\frac{\left(\boldsymbol{\Omega}\times\mathbf{R_{\star}}\right)\times\mathbf{B}}{c},
\label{eq_efield}
\end{equation}
where $R_{\star}=r_{\star}\sin\theta$. On the Yee-mesh, only $E_{\theta}$ is fixed by Eq.~(\ref{eq_efield}), while $E_{\phi}=0$ at all times. The fields at the outer radial boundary are dampened by a spherical shell of absorbing material located between $r_{\rm abs}=9 r_{\star}=3R_{\rm LC}$ and $r_{\rm max}$ (Figure~\ref{fig_geo}, see \citealt{2015MNRAS.448..606C} for more details). On the rotation axis ($\theta=0,~\pi$) we enforce $\partial E_{\rm r}/\partial\theta=0$, $E_{\rm \phi}=0$ and $B_{\rm \theta}=0$ (see \citealt{1983ITNS...30.4592H} for an alternative implementation for finding $E_{\rm r}$). Along the $\phi$-direction, the standard periodic boundary condition is applied to all fields.

For the particles, we use the same injection procedure as in \citet{2015MNRAS.448..606C}, where a large plasma supply is launched from the stellar surface which then fills the magnetosphere entirely, and hence, provides the quasi force-free configuration we are seeking for this study. However, it ignores the details of the pair creation physics which is the main focus of other studies \citep{2013MNRAS.429...20T, 2014ApJ...795L..22C, 2015ApJ...801L..19P}. At each time step, the star injects uniformly in $\theta$ and $\phi$ twice the fiducial Goldreich-Julian plasma density, defined as $n^{\star}_{\rm GJ}\equiv \Omega B_{\star}/2\pi e c$ where $e$ is the electron charge, in the form of electron-positron pairs. To allow the plasma to escape and populate the magnetosphere, each pair is generated with an initial poloidal velocity along the field lines, $v_{\rm pol}=0.5c$. The particles are also created in co-rotation with the star. To avoid the accumulation of a large plasma density close to the star (in particular in the region of close field lines), the code stops injecting new particles if the fiducial plasma multiplicity $\kappa_{\star}\equiv n_{\star}/n^{\star}_{\rm GJ}$ exceeds $10$, where $n_{\star}$ is the plasma density at $r=r_{\star}$. Particles are removed from the simulation if they hit the star or if they reach the absorbing layer ($r>r_{\rm abs}$). Particles are excluded from the rotation axis by bouncing off specularly. In the azimuthal direction, periodic boundary conditions apply. The results are not very sensitive to the choice of boundary conditions applied to the particles along the axis because there is not significant current and energy outflow in this region.

In this work, we ran a series of 7 simulations where the obliquity angle varies in the range $\chi=0^{\rm o},$ $15^{\rm o},$ $30^{\rm o},$ $45^{\rm o},$ $60^{\rm o},$ $75^{\rm o},$ $90^{\rm o}$. The box is composed of $1024\times256\times256$ grid cells in $r$, $\theta$, $\phi$ respectively. The electron collisionless skin depth $d_{\rm e}$ is resolved everywhere by at least 2 cells, the lower bound corresponds to $\kappa_{\star}=10$ at $r=r_{\star}$. In the current sheet close to the light-cylinder, the typical particle gyro-radius is well resolved by about $30$ cells. The plasma frequency is $\Delta t\lesssim 0.032\omega^{-1}_{\rm pe}$. In fact, the time step was chosen to resolve the smallest particle gyro-frequency in the simulation $\omega^{\star}_{\rm L}=e B_{\star}/\gamma_{\star} m_{\rm e} c^2$, i.e., $\Delta t\omega^{\star}_{\rm L}\approx 1$ (where $m_{\rm e}$ is the electron mass and $\gamma_{\star}\approx 1$ is the particle Lorentz factor at the surface). We found that this condition must be fulfilled to recover the correct radiative energy losses by the particles (see Sect.~\ref{sect_frad} below). Note that the corresponding Larmor radius is not resolved by the grid because of our limited computational power. The largest timescale of the problem is the pulsar spin period, $P$, which is reached after about $1.2\times10^5$ time steps. The simulations ran until $t=2P$, although the solutions approach a quasi-steady state after one spin period only. Once the magnetosphere is established everywhere, the total number of macro-particles is of order $\sim 2\times 10^8$ which gives an average of about $3$ particles per cell, with a higher concentration in the current sheet and in the region of closed field lines. The fiducial magnetization parameter at the surface of the star, defined as $\sigma_{\star}\equiv B^2_{\star}/4\pi\kappa_{\star}n^{\star}_{\rm GJ} m_{\rm e} c^2$, is $\sigma_{\star}=500$. The magnetization parameter estimated at the light cylinder and just above the current sheet is $\sigma_{\rm LC}= B^2_{\rm LC}/4\pi n_{\rm LC} \Gamma_{\rm LC}m_{\rm e} c^2\approx 50$ ($n_{\rm LC}$ and $\Gamma_{\rm LC}\approx 2$ are respectively the plasma density and the wind Lorentz factor at the light cylinder). Table~\ref{table1} summarizes the list of the physical and numerical parameters employed in this study. Realistic values cannot be set for all the physical parameters due to the current limits in computing power, but the microscopic and the macroscopic scales are well-separated by several orders of magnitude.

\begin{table}
 \caption{List of the physical and numerical parameters as defined in the text, and their values used in this study. In this table, $r_{\star}$ refers to the radius of the neutron star, $\Delta r$ is the radial grid spacing, and $\Delta t$ is the simulation time step.}
 \label{table1}
 \begin{tabular}{lcc}
  \hline
  Name & Symbol & Values \\
  \hline
  Obliquity & $\chi$ & $0^{\rm o},15^{\rm o},30^{\rm o},45^{\rm o},60^{\rm o},75^{\rm o},90^{\rm o}$ \\
  \# grid cells & $N_{\rm r}\times N_{\theta}\times N_{\phi}$ & $1024\times256\times256$\\
  \# particles & $N_{\rm tot}$ & $\sim 2\times 10^8$\\
  Inner radius & $r_{\rm min}/r_{\star}$ & $1$\\
  Light cylinder & $R_{\rm LC}/r_{\star}$ & $3$\\
  Absorb radius & $r_{\rm abs}/r_{\star}$ & $9$\\
  Outer radius & $r_{\rm max}/r_{\star}$ & $10$\\
  Range in $\theta$ & $\theta_{\rm min},\theta_{\rm max}$ & $0,\pi$\\
  Range in $\phi$ & $\phi_{\rm min},\phi_{\rm max}$ & $0,2\pi$\\
  Skin-depth & $d_{\rm e}/\Delta r$ & $>2$\\
  Plasma freq. & $\omega^{-1}_{\rm pe}/\Delta t$ & $>31$\\
  Larmor freq. & $\omega^{-1}_{\rm L}/\Delta t$ & $>1$\\
  Pulsar period & $P/\Delta t$ & $1.2\times 10^5$\\
  Sync. time & $t^{\star}_{\rm sync}/\Delta t$ & $6$\\
  Magnetization & $\sigma_{\star}$ & $500$\\
  Mag. at LC & $\sigma_{\rm LC}$ & $50$\\
  \hline
 \end{tabular}
\end{table}
 
\section{Modeling curvature and synchrotron radiation}\label{sect_rad}

\subsection{Radiation reaction force}\label{sect_frad}

The equation that governs the motion of a particle subject to radiative energy losses is the Abraham-Lorentz-Dirac equation, i.e.,
\begin{equation}
\frac{d(\gamma m_{\rm e} \mathbf{v})}{dt}=q\left(\mathbf{E}+\boldsymbol{\beta}\times\mathbf{B}\right)+\mathbf{g},
\label{eq_motion}
\end{equation}
where $\mathbf{v}=\boldsymbol{\beta}c$ is the particle 3-velocity, $\gamma=1/\sqrt{1-\beta^2}$ is the particle Lorentz factor, and $q$ is the particle electric charge. The first terms on the right-hand side in Eq.~(\ref{eq_motion}) is the usual Lorentz force, while $\mathbf{g}$ is the radiation reaction force due to the emission of photons by the accelerated particle (here curvature and synchrotron radiation). This force must be added in the PIC code because the frequency of the radiation emitted by relativistic particles (i.e., $\gamma\gg1$) is not resolved by the grid, and hence the back-reaction on the particle motion cannot be captured. Within the framework of classical electrodynamics, the radiation reaction force is given by the Landau-Lifshitz formula
\begin{eqnarray}
\mathbf{g} =\frac{2}{3}r^2_{\rm e}\left[\left(\mathbf{E}+\boldsymbol{\beta}\times\mathbf{B}\right)\times\mathbf{B}+\left(\boldsymbol{\beta}\cdot\mathbf{E}\right)\mathbf{E}\right]\hspace{2cm}\nonumber \\
-\frac{2}{3}r^2_{\rm e}\gamma^2\left[\left(\mathbf{E}+\boldsymbol{\beta}\times\mathbf{B}\right)^2-\left(\boldsymbol{\beta}\cdot\mathbf{E}\right)^2\right]\boldsymbol{\beta},
\label{frad}
\end{eqnarray}
where $r_{\rm e}=e^2/m_{\rm e}c^2$ is the classical radius of the electron. In this expression, we have intentionally omitted the term that contains the total time derivative of the fields (i.e., $\partial/\partial t+\mathbf{v}\cdot\mathbf{\nabla}$), which is negligible compared to the two terms reported here \citep{2010NJPh...12l3005T}. For ultra-relativistic particles, the term proportional to $\gamma^2$ is clearly dominant, and corresponds to a drag force opposite to the particle velocity that is proportional to the emitted radiative power.

However, we found that the first term must be included in the code, even if $\gamma\gg 1$, in order to capture the correct curvature radiation cooling rate, i.e., $P_{\rm curv}\propto \gamma^4$. Indeed, a particle moving along a curved field line is subject to a centrifugal force and, therefore, it drifts perpendicular to the plane of curvature. The curvature drift velocity divided by $c$ is
\begin{equation}
\beta_{\rm cd}=\frac{\gamma m_{\rm e} c^2}{e B R_{\rm c}},
\end{equation}
which corresponds to the ratio of the relativistic particle Larmor radius, $R_{\rm L}=\gamma m_{\rm e} c^2/eB$, to the radius of curvature, $R_{\rm c}$. In pulsars, we have $R_{\rm L}\ll R_{\rm c}$ so that the drift velocity is non-relativistic, i.e., $\beta_{\rm cd}\ll 1$. Along the curvature drift direction, the ratio of the first term over the second term is $1/\gamma^2\beta^2_{\rm cd}$. This ratio indicates that if $\gamma < R_{\rm c}/R_{\rm L}$, the first term dominates, even if $\gamma\gg 1$, and gives the correct equilibrium velocity in this direction, which then enters in the drag term. This condition is fulfilled in both the simulations and in real pulsars. To summarize, both terms in Eq.~(\ref{frad}) are necessary to recover the correct curvature radiation reaction force. As already mentioned in Sect.~\ref{sect_setup}, another condition must be fulfilled: the Larmor frequency $\omega^{\star}_{\rm L}$ has to be resolved by the simulation. Neglecting the non-relativistic term or under-resolving $\omega^{\star}_{\rm L}$ would lead to an overestimation of the power lost by the particle.

The strength of the radiation reaction force is amplified by a constant numerical factor, $\kappa_{\rm rad}$, such that in the code units, the fiducial synchrotron cooling time of a $\gamma=1$ particle at the surface of the star is as short as possible but also well resolved by the simulation, $t^{\star}_{\rm sync}\equiv -\gamma/\dot{\gamma}_{\rm sync}=9 m_{\rm e}c/4\kappa_{\rm rad}r^2_{\rm e}B^2_{\star}\approx 6\Delta t$. At the light-cylinder, the typical particle cooling time is $t^{\rm LC}_{\rm sync}\approx 85\Delta t$.

\subsection{Radiation spectra}

The radiation power spectrum emitted by a single relativistic particle is given by (e.g., \citealt{1970RvMP...42..237B})
\begin{equation}
F_{\nu}\left(\nu\right)=\frac{\sqrt{3}e^3 \tilde{B}_{\perp}}{m_{\rm e}c^2}\left(\frac{\nu}{\nu_{\rm c}}\right)\int_{\nu/\nu_{\rm c}}^{+\infty}K_{5/3}(x)dx,
\label{eq_flux}
\end{equation}
where $\nu$ is the radiation frequency, $K_{5/3}$ is the modified Bessel function of $5/3$ order,
\begin{equation}
\tilde{B}_{\perp}=\sqrt{\left(\mathbf{E}+\boldsymbol{\beta}\times\mathbf{B}\right)^2-\left(\boldsymbol{\beta}\cdot\mathbf{E}\right)^2},
\label{eq_bperp}
\end{equation}
and
\begin{equation}
\nu_{\rm c}=\frac{3 e \tilde{B}_{\perp}\gamma^2}{4\pi m_{\rm e}c}
\label{eq_nuc}
\end{equation}
is the critical frequency. The above expressions are identical to the usual synchrotron formulae, but they are more general in the sense that they are valid for both synchrotron and curvature radiation, or for a mix of both (synchro-curvature radiation, see \citealt{1996ApJ...463..271C, 2013arXiv1305.0783P, 2015AJ....149...33K, 2015MNRAS.447.1164V}). The only difference with the classical formulae is shown in Eq.~(\ref{eq_bperp}). In the synchrotron regime, $\tilde{B}_{\perp}$ can be interpreted as the relativistically invariant magnetic field strength perpendicular to the particle motion, instead of just $B_{\perp}$ as it is in the usual synchrotron expressions. If curvature radiation dominates, this quantity is related to the local radius of curvature of the field lines, such that $R_{\rm c}=\gamma m_{\rm e} c^2/e\tilde{B}_{\perp}$ \citep{2015AJ....149...33K}. Note that there is no need to compute the curvature from the global structure of the field lines, only local quantities suffice which greatly simplifies the numerical procedure. In practice, {\tt ZELTRON} computes and stores the value of $\tilde{B}_{\perp}$ ($E$ and $B$ are interpolated at the particle position from the grid) at every time step, which is then used to compute the emitted radiation spectrum per particle.

\begin{figure}
\centering
\includegraphics[width=8cm]{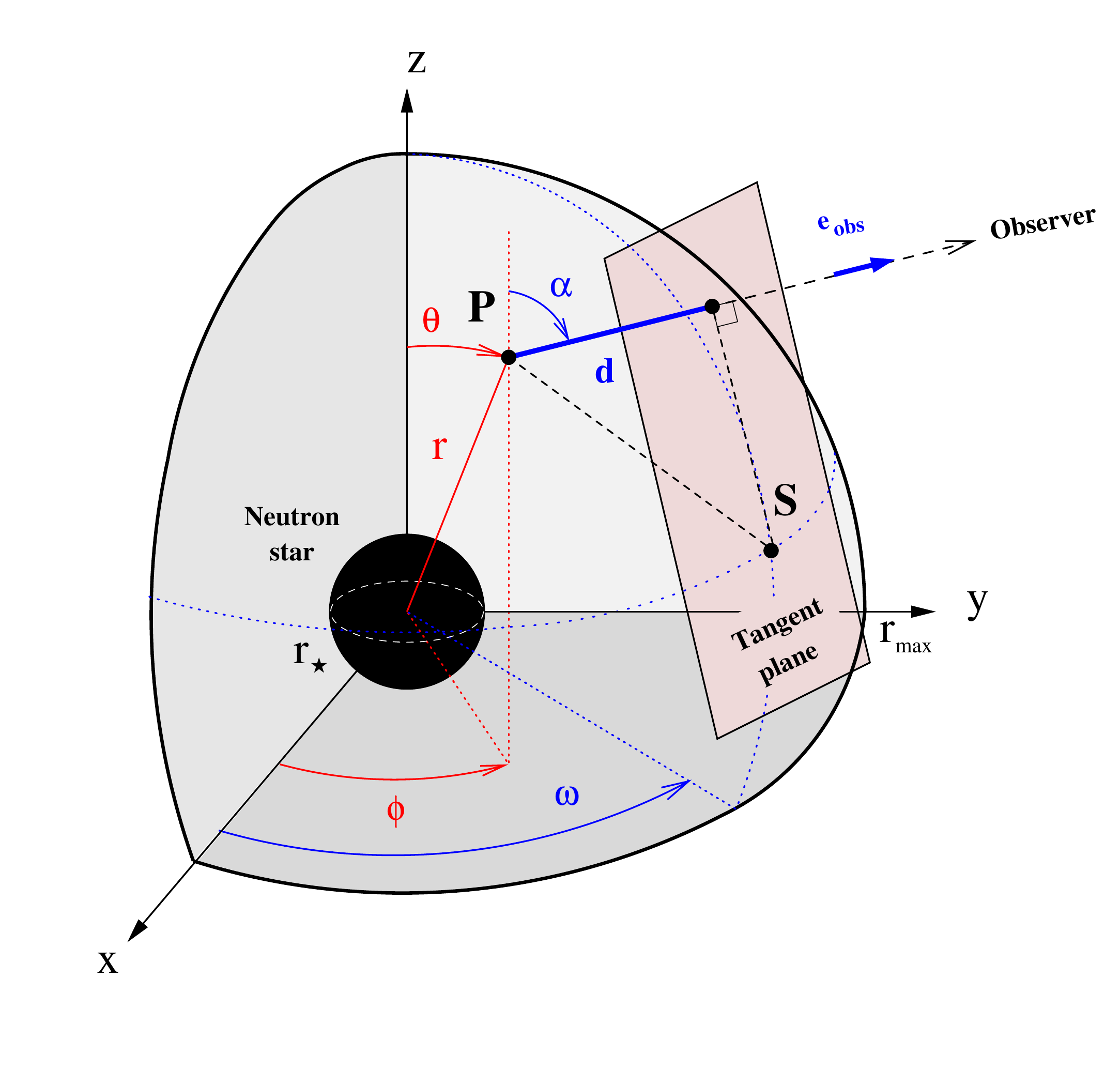}
\caption{This diagram presents the geometrical quantities used here to compute lightcurves. A photon is emitted by a particle located at the point P $(r,\theta,\phi)$ towards the observer, along the unit vector $\mathbf{e_{\rm obs}}$ (co-latitude $\alpha$, azimuth $\omega$). The photon will be received by the observer after a time delay $t_{\rm d}={\rm d}/c=\left(\mathbf{PS}\cdot\mathbf{e_{\rm obs}}\right)/c$ with respect to the closest point to the observer, the point S $(r_{\rm max},\alpha,\omega)$.}
\label{fig_geo_light}
\end{figure}

\subsection{Lightcurves}\label{sect_light}

Now that we know how each particle moves and radiates, we have all the elements at hand to compute the high-energy pulsar lightcurves, directly from the PIC simulations. The modeling of the full radiative transfer in pulsar magnetospheres is a difficult problem by itself, so we propose to use the following simplifying assumptions. Once emitted, the photons cannot be absorbed by the magnetic field or by other photons, unless they hit the star. In such a case they are removed from the simulation. Photons propagate freely everywhere in the magnetosphere on straight lines at the speed of light. All photons are beamed along the emitting particle's direction of motion. This assumption is valid only for ultra-relativistic particles, for which the emission is focused within a cone of semi-aperture angle $\sim 1/\gamma\ll 1$. Radiative processes other than curvature and synchrotron are neglected in this study (i.e., inverse Compton, synchrotron self-Compton, bremsstrahlung). In {\tt ZELTRON}, each macro-particle radiates a single ``macro-photon'' (or simply ``photon'' in the following) in the simulation frame. Each macro-photon represents a bunch of physical photons with the power spectrum given in Eq.~(\ref{eq_flux}). The (macro-)photons are then collected on a spherical screen located at $r=r_{\rm max}$, where the radiation flux is reconstructed as a function of the viewing angle $\alpha$ (co-latitude) and $\omega$ (azimuth), and the radiation frequency $\nu$. 

To build lightcurves, we need to account for the phase shift due to the finite propagation time of the photons to the observer. To do this, consider a particle located at the point P of coordinates $(r,\theta,\phi)$ at time $t$ that radiates photons along the direction $(\alpha,\omega)$ shown by the unit vector $\mathbf{e_{\rm obs}}$ in Figure~{\ref{fig_geo_light}}. Then, the time delay relative to the point closest to the observer, namely the point S of coordinates  $(r_{\rm max},\alpha,\omega)$, is given by the shortest distance to the plane tangent to the sphere of radius $r=r_{\rm max}$ and passing through S (the length ${\rm d}$ in Figure~\ref{fig_geo_light}), divided by $c$, i.e.,
\begin{equation}
t_{\rm d}=\frac{\left(\mathbf{PS}\cdot\mathbf{e_{\rm obs}}\right)}{c},
\end{equation}
with
\begin{equation}
\left(\mathbf{PS}\cdot\mathbf{e_{\rm obs}}\right)=r_{\rm max}-r\left[\sin\alpha\sin\theta\cos\left(\omega-\phi\right)+\cos\alpha\cos\theta\right].
\end{equation}
Hence, the photons arrive at the pulsar phase
\begin{equation}
\Phi_{\rm P}\equiv\frac{1}{2\pi}{\rm Modulo}\left[\omega-\Omega t_{\rm d},2\pi\right].
\end{equation}
The phase $\Phi_{\rm P}=0$ is in the plane containing $\boldsymbol{\mu}$ and $\boldsymbol{\Omega}$.

\section{results}\label{sect_results}

After about one spin period, the pulsar magnetosphere has already reached a quasi-steady, quasi-force-free configuration characterized by abundant plasma everywhere, and a prominent undulating current sheet beyond the light cylinder. The pulsar spindown measured at the light cylinder, $L_{\rm P}$, is compatible with the force-free solution \citep{2006ApJ...648L..51S} and previous PIC simulations \citep{2015ApJ...801L..19P}, see Table.~\ref{tab_spindown}. The presence of the radiation reaction force in the simulation does not affect the overall structure of the magnetosphere\footnote{However, we note that the current layer thickness decreases with strong radiative cooling \citep{2011PhPl...18d2105U}.} (i.e., current distribution, field morphology, spindown), because the basic picture of particles moving along the field lines at the speed of light holds with or without radiative cooling (the current depends only on the particle 3-velocity). This is why we will not expand further the discussion on the structure of the oblique pulsar magnetosphere here (see the previous study by \citealt{2015ApJ...801L..19P} for more details), and instead focus on particle acceleration and radiation in the magnetosphere.

\subsection{Spatial distribution of energetic particles and radiation}

\begin{figure*}
\centering
\includegraphics[width=17cm]{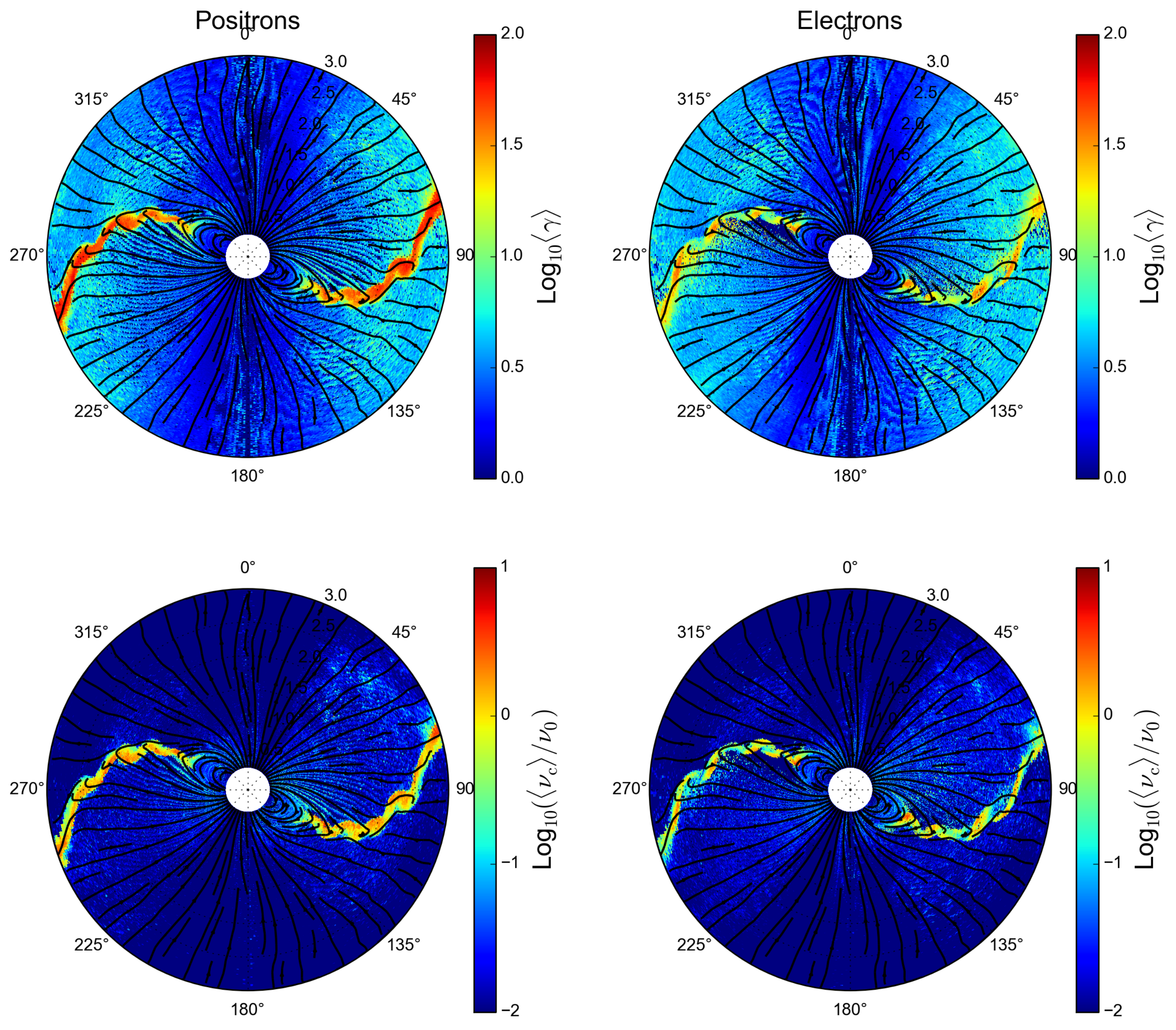}
\caption{Cell-averaged particle Lorentz factor (top left: positrons, top right: electrons), $\langle\gamma\rangle$, and radiation critical frequency (bottom left: emitted by positrons, bottom right: emitted by electrons), $\langle\nu_{\rm c}\rangle$ normalized to the fiducial synchrotron frequency $\nu_0\equiv 3 e B_{\star}/4\pi m_{\rm e} c$, for a pulsar obliquity $\chi=30^{\rm o}$ at time $t=1P$ in the $(\boldsymbol{\Omega},\boldsymbol{\mu})$-plane. The domain ranges from the surface of the star $r_{\rm min}=r_{\star}$ to $r_{\rm abs}=9r_{\star}=3R_{\rm LC}$, distances are in units of the light-cylinder radius, $R_{\rm LC}=3r_{\star}$. The poloidal magnetic field lines are shown by the black solid lines. The central white disk is the neutron star.}
\label{fig_energy}
\end{figure*}

The top panels in Figure~\ref{fig_energy} shows the positron (left) and electron (right) Lorentz factors averaged in each cell, $\langle\gamma\rangle$, for $\chi=30^{\rm o}$ in the ($\boldsymbol{\Omega},\boldsymbol{\mu}$)-plane at $t=1P$. For both species, the most energetic particles are located within the equatorial current sheet beyond the light-cylinder radius and the Y-point (i.e., at the base of the sheet at $r\approx R_{\rm LC}$, see e.g. \citealt{2003ApJ...598..446U}), where reconnection takes place and accelerates particles. This result is compatible with previous studies of the aligned pulsar \citep{2014ApJ...785L..33P, 2015MNRAS.448..606C}. However, we note that energetic electrons are not found along the separatrix layers as in \citet{2015MNRAS.448..606C}, because they cool abruptly at the Y-point on their way from the current sheet back to the star as they feel a sharp increase in $\tilde{B}_{\perp}$. As in the aligned case \citep{2014ApJ...785L..33P, 2015MNRAS.448..606C}, the particle Lorentz factor in the current sheet is given by the magnetization parameter at the light-cylinder $\sigma_{\rm LC}\approx 50$. This result still holds in the strong radiative cooling regime explored here, because deep in the current sheet, the effective perpendicular magnetic field $\tilde{B}_{\perp}$ is small and hence their energy is limited by the total available magnetic energy per particle (i.e., $\sigma_{\rm LC}$) rather than limited by radiative cooling \citep{2004PhRvL..92r1101K, 2007A&A...472..219C, 2013ApJ...770..147C}. Away from the equatorial regions ($\theta<\pi/2-\chi$ and $\theta>\pi/2+\chi$), the particles fly radially outward at approximatively the $\mathbf{E}\times\mathbf{B}$ drift velocity and thus accelerate slowly with the cylindrical radius $R=r\sin\theta$ to form a mildly relativistic ($\Gamma\approx 2$-$3$) pulsar wind \citep{2015MNRAS.448..606C}. Within the equatorial regions ($\pi/2-\chi<\theta<\pi/2+\chi$), the wind particles are more energetic than in the polar regions with $\langle\gamma\rangle\sim 10$, in particular close to the current sheet. This acceleration could be connected to the reconnection induced inflow towards the sheet \citep{2013MNRAS.435L...1T}. It is also more efficient at higher inclinations as a larger portion of the wind feeds the current layer from the polar cap directly. Inside the light cylinder, we do not find significant particle acceleration, $\langle\gamma\rangle<10$.

\begin{figure}
\centering
\includegraphics[width=8.6cm]{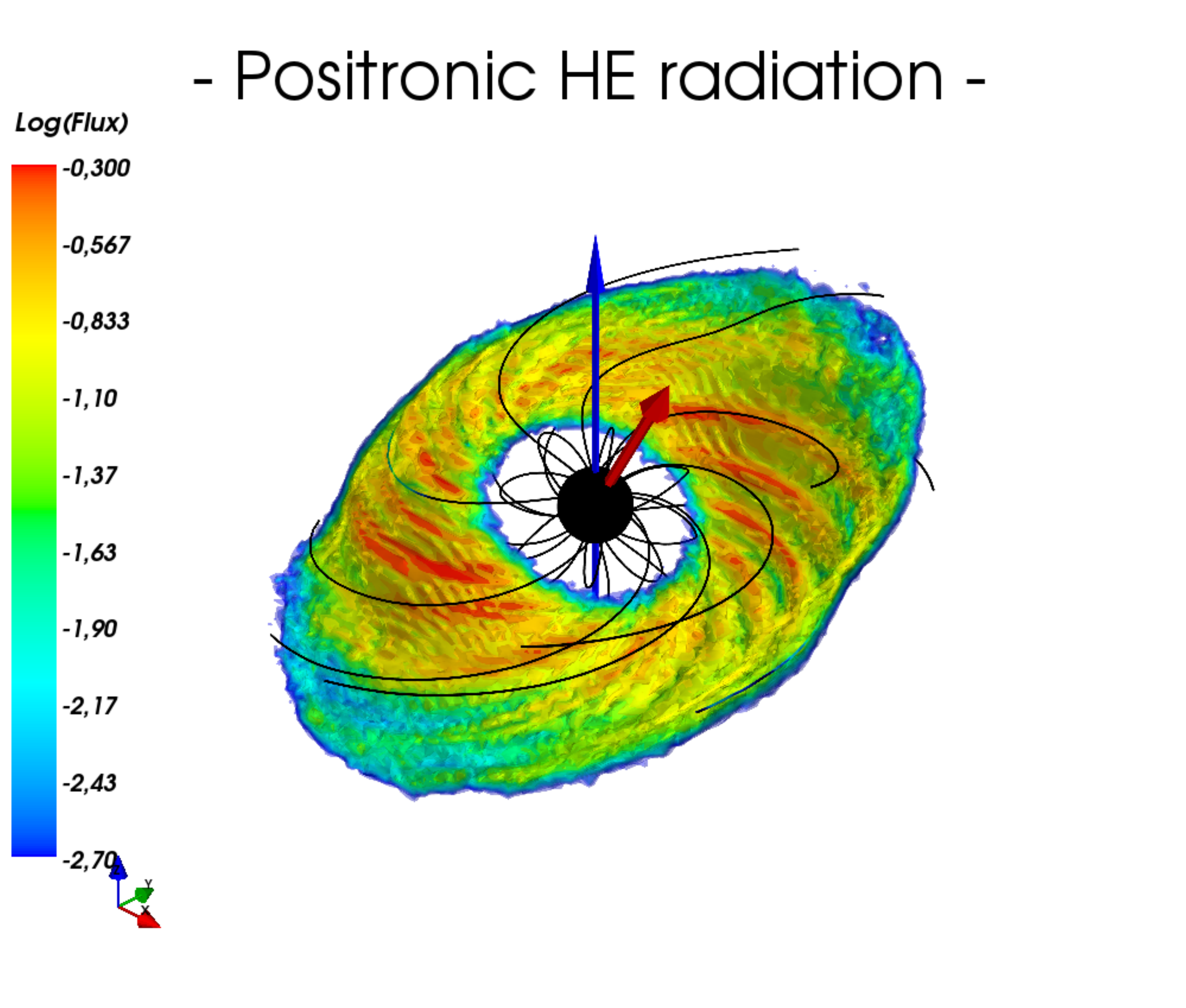}
\includegraphics[width=8.6cm]{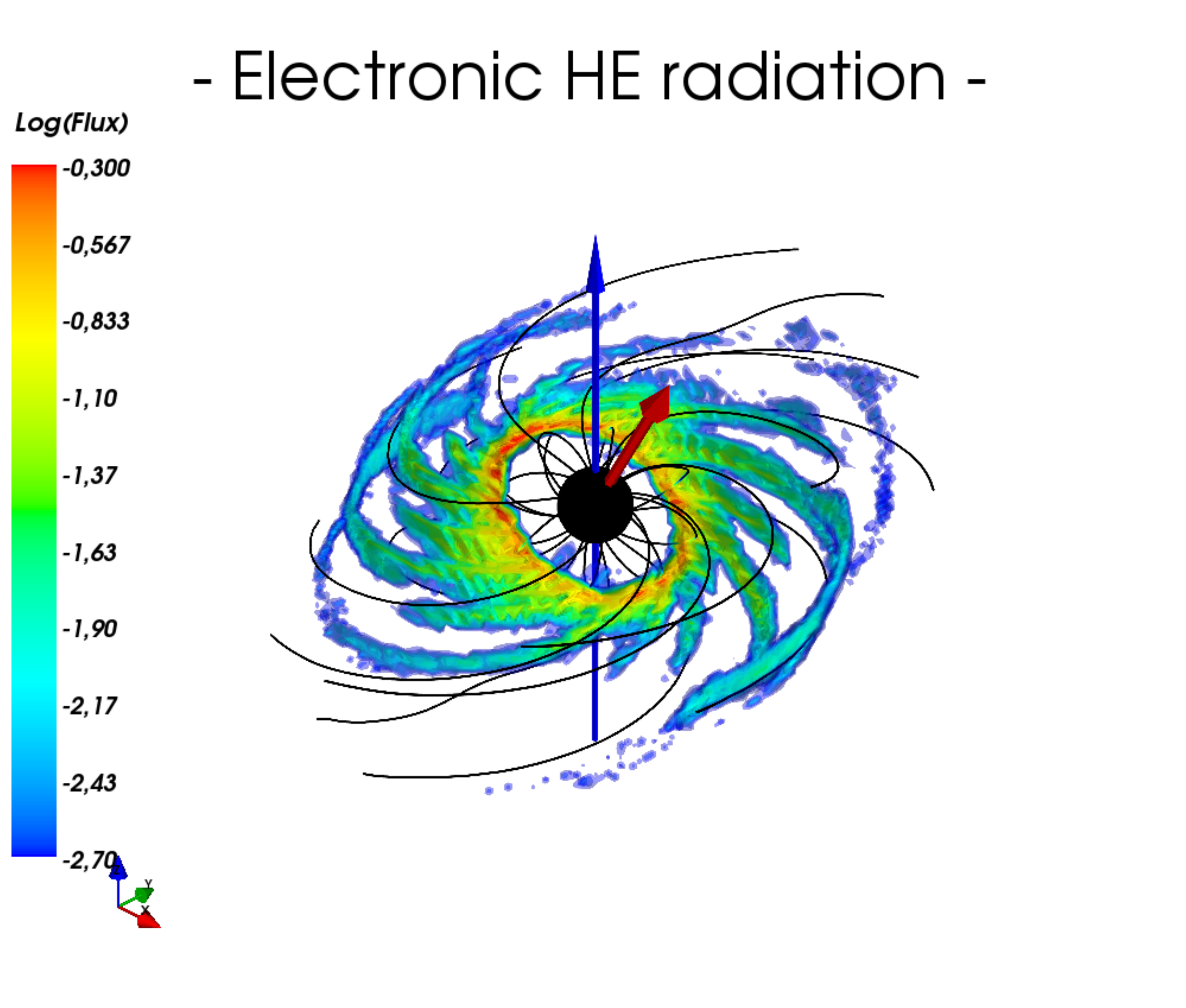}
\caption{Three-dimensional rendering of the total radiation flux integrated above $\nu_0$ (colors are in logarithmic scale, Log$_{10}F_{\nu}(\nu>\nu_0)$) and from all directions, emitted by the positrons (top) and the electrons (bottom). The black solid lines show the last closed field lines confined within the light-cylinder, and the first open field lines located just above and below the current sheet. The arrows are along the magnetic (red) and the rotation (blue) axis of the pulsar (the black sphere) for $\chi=30^{\rm o}$. The radius varies from $r_{\star}$ to $3R_{\rm LC}$.} 
\label{fig3d_flux}
\end{figure}

\begin{figure}
\centering
\includegraphics[width=8.6cm]{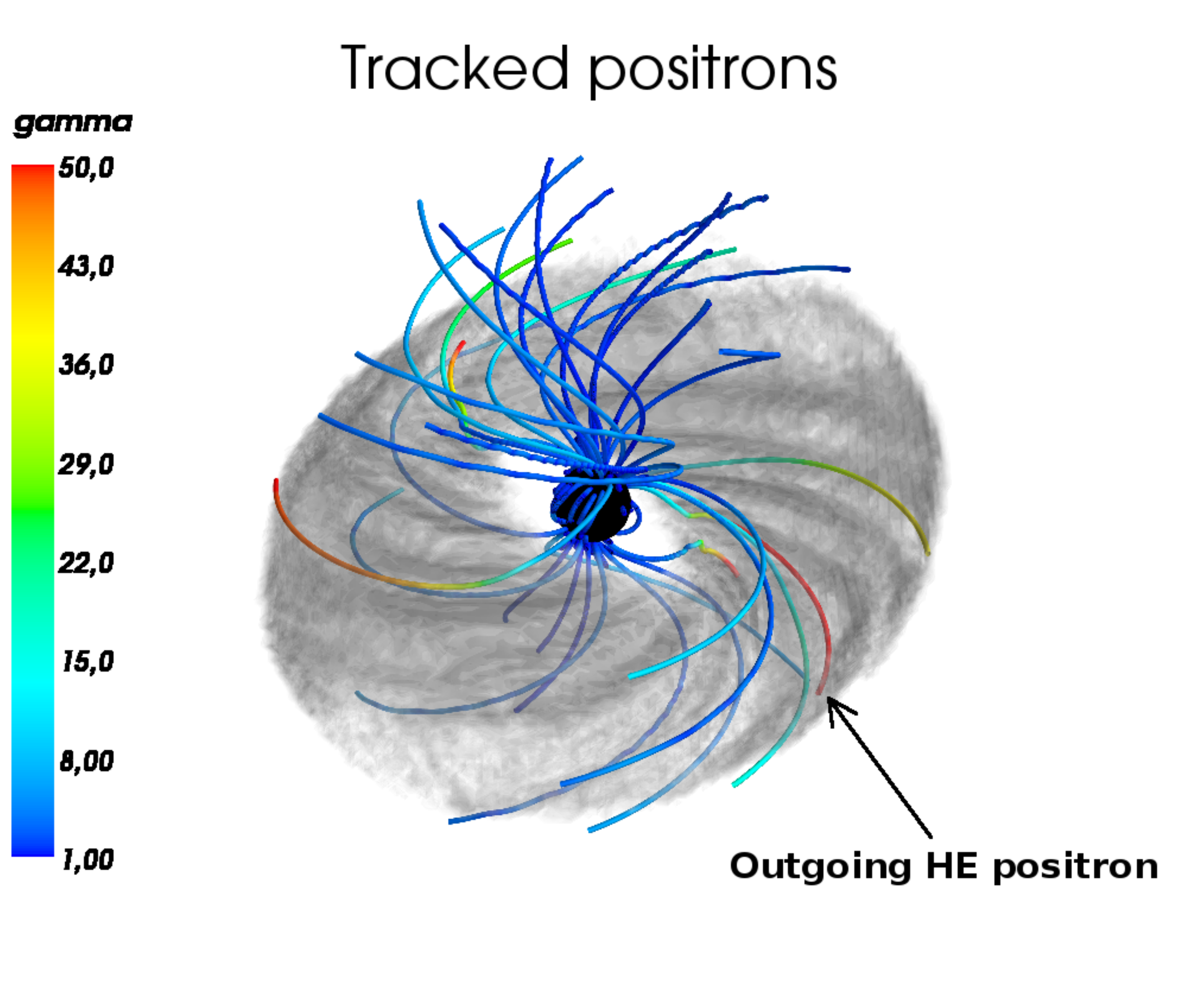}
\includegraphics[width=8.6cm]{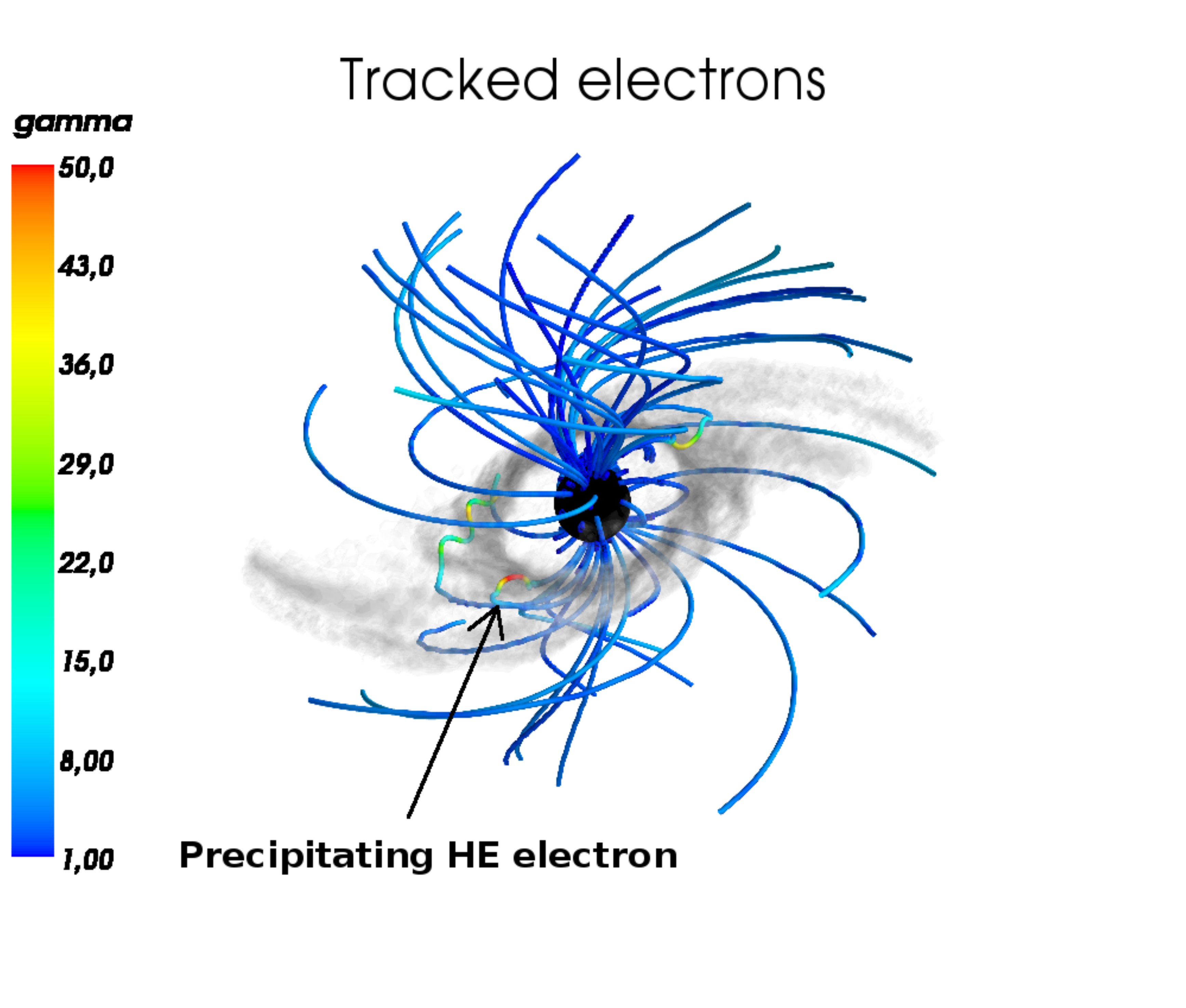}
\caption{Three-dimensional trajectories of a randomly selected sample of 98 positrons (top) and 105 electrons (bottom), followed from the time of their injection at $t=1 P$ at the surface of the star until $t=2P$, for $\chi=30^{\rm o}$. The color shows the Lorentz factor of the particles along their trajectories. The trajectories are drawn in the co-rotating frame to show that electrons and positrons are accelerated within the equatorial current sheet. In the lab frame, particle trajectories are almost straight radial lines. The high-energy radiation flux presented in Figure~\ref{fig3d_flux} is reported here in black and white. The viewing angle is different in each figure to have a clear view of the few high-energy particle trajectories contained in this sample. Magnetic fields lines are omitted for clarity.} 
\label{fig3d_orbits}
\end{figure}

The bottom panel in Figure~\ref{fig_energy} presents the cell-averaged critical radiation frequency $\langle\nu_{\rm c}\rangle$ (as defined in Eq.~\ref{eq_nuc}) emitted by positrons (left) and electrons (right). Frequencies are normalized by $\nu_0$ which corresponds to the synchrotron frequency of a $\gamma=1$ particle immersed in $\tilde{B}_{\perp}=B_{\star}$, i.e., $\nu_0\equiv 3 e B_{\star}/4\pi m_{\rm e} c$ which corresponds to $\nu_0\approx 40~$MeV in the code units. The resulting maps are nearly identical to the particle energy maps. Hence, this indicates that the high-energy photons ($\nu\geq\nu_0$) are emitted by the high-energy particles accelerated in the current sheet via synchrotron radiation, rather than curvature radiation inside the light-cylinder. To illustrate this, we display in Figure~\ref{fig3d_flux} the spatial distribution of the total radiation flux integrated above $\nu_0$, emitted by the positrons (top) and the electrons (bottom). The 3D rendering of the energetic radiation matches exactly with the equatorial current distribution. The analysis of individual particle trajectories shows that the high-energy positrons radiate continuously while moving away from the pulsar, within the current sheet (see top panel in Figure~\ref{fig3d_orbits}). In contrast, energetic electrons radiate mostly at the base of the current sheet (i.e., close to the Y-point which looks more like a bright ring in 3D) as they precipitate back towards the star (see bottom panel in Figure~\ref{fig3d_orbits}). Figure~\ref{fig3d_flux} highlights also the kinked spiral structure of the current sheet due to plasma instabilities (kink and tearing modes). The tearing instability creates plasma over-densities that translate into brighter spiral arms in the current sheet. While the tearing instability seems active at all inclinations, we observe that the strength of the kink instability decreases with pulsar obliquity, in agreement with \citet{2015ApJ...801L..19P}.

The equatorial wind is emitting at intermediate frequencies for $\chi=30-45^{\rm o}$, $\langle\nu_{\rm c}\rangle\sim 0.1 \nu_0$, and hence does not appear in Figure~\ref{fig3d_flux}, but radiates at frequencies up to $\langle\nu_{\rm c}\rangle\gtrsim\nu_0$ for $\chi=90^{\rm o}$, i.e., comparable to the photons from the sheet. In contrast, the particles in the polar wind outside $R_{\rm LC}$ have $\langle\nu_{\rm c}\rangle/\nu_0\ll1$ because $\tilde{B}_{\perp}\approx 0$ there. Indeed, if the particle velocity coincides with the $\boldsymbol \beta =\mathbf{E} \times \mathbf{B}/ \mathbf{B}^2$ drift velocity and if $\mathbf{E} \cdot \mathbf{B}=0$, then according to Eq.~(\ref{eq_bperp}) $\tilde{B}_{\perp}$ vanishes. Within the light cylinder, a mix of curvature and synchrotron radiation is emitted by both species with a typical frequency $\nu\sim 0.1\nu_0$. Most of the low-energy radiation is emitted by the mildly relativistic particles injected at the surface with $\gamma\approx 1.25$. Note that there is no energetic radiation coming from $R<R_{\rm LC}$ in Figure~\ref{fig3d_flux}, since all parallel electric field is efficiently screened inside the light cylinder with the plasma injection used in this study. This conclusion is robust against inclination.

\begin{table}
 \caption{Pulsar spindown, $L_{\rm P}$, and high-energy radiative efficiency, $\eta_{\gamma}$ (see definition in Sect.~\ref{sect_spec}), as a function of the obliquity $\chi$, in units of $L_0=\mu^2\Omega^4/c^3$.}
 \label{tab_spindown}
 \begin{tabular}{lccccccc}
  \hline
  $\chi$ & $0^{\rm o}$ & $15^{\rm o}$ & $30^{\rm o}$ & $45^{\rm o}$ & $60^{\rm o}$ & $75^{\rm o}$ & $90^{\rm o}$ \\
  \hline
  $L_{\rm p}/L_0$ & $0.96$ & $1.04$ & $1.25$ & $1.51$ & $1.74$ & $1.91$ & $1.95$ \\
  \hline
  $\eta_{\gamma}$ & $9.1\%$ & $6.9\%$ & $4.0\%$ & $2.2\%$ & $1.7\%$ & $1.7\%$ & $1.8\%$ \\
  \hline
 \end{tabular}
\end{table}

\subsection{Total particle and radiation spectra}\label{sect_spec}

Figure~\ref{fig_spectra} shows the particle ($\gamma dN/d\gamma$) and radiation spectral energy distributions ($\nu F_{\nu}$ using Eq.~\ref{eq_flux}), averaged over the whole box and over all directions. These are not the observed spectra, because the particle and radiation angular distributions are highly anisotropic. The observed phase-averaged spectra are discussed in the next section. At low inclinations $\chi\lesssim 45^{\rm o}$, the particle and photon spectra present two clear components: (i) low-energy particles $(\gamma\approx$ a few) radiating synchro-curvature radiation with $\nu\sim 0.1\nu_0$ inside the light-cylinder, and (ii) quasi-monoenergetic high-energy particles $(\gamma\gtrsim \sigma_{\rm LC}\approx 50)$ emitting synchrotron radiation in the equatorial current sheet with $\nu\sim 5\nu_0$ followed by an exponential cutoff. There is also a clear excess of energetic positrons, about 10 times more than energetic electrons, as found in the aligned case \citep{2015MNRAS.448..606C}. In contrast, at higher inclinations ($\chi\gtrsim 45^{\rm o}$) there is no clear separation between the low- and the high-energy particles, and both species present nearly identical spectra for $\chi\rightarrow 90^{\rm o}$, as it should by symmetry. The high-energy synchrotron component from the current sheet is much less prominent in this case. In real pulsars, the separation of scale between the low- and the high-energy particles is more pronounced than what is achieved in the simulations. This ratio could be as high as about $10^4$ in Crab-like pulsars, which has to be compared with about $10^2$ here. Hence, the contribution from the sheet and the low-energy particles from the magnetosphere should be well-separated even for high-obliquity pulsars.

From this figure, we can infer how much of the pulsar spindown power is channeled into energetic radiation. We define the high-energy radiative efficiency, $\eta_{\rm \gamma}$, as the frequency-integrated radiative power above $\nu_0$ divided by the pulsar spindown $L_{\rm P}$, i.e., $\eta_{\gamma}=\frac{1}{L_{\rm P}}\int_{\nu_0}^{+\infty}\kappa_{\rm rad}F_{\nu}d\nu$. The radiative efficiencies obtained in these simulations are reported in Table.~\ref{tab_spindown}. We observe a clear trend of decreasing radiative efficiency with pulsar obliquity, starting from about $9\%$ for $\chi=0^{\rm o}$ to about $1-2\%$ for $\chi\gtrsim 60^{\rm o}$. Here again, these numbers do not correspond to the observed (apparent) efficiencies which depend on the observer's viewing angle as discussed below.

\subsection{Observed high-energy emission}

Now that we have identified the locations and the nature of the radiation emitted in the simulations, we can focus on the radiative output as seen by a distant observer, taking into account finite light-crossing time (see Sect.~\ref{sect_light}). Figure~\ref{fig_sky_freq} presents the radiation flux emitted by positrons only, as function of the observer's viewing angle $\alpha$ and pulsar phase $\Phi_{\rm P}$ (i.e., the ``skymap''), in the low-energy band ($\nu<\nu_0$, top panel) and the high-energy band ($\nu>\nu_0$, bottom panel) for $\chi=30^{\rm o}$. The low-energy skymap is characterized by two bright features separated by $0.5$ in phase and whose centroid is about $\chi=\pm 30^{\rm o}$ away from the poles ($\alpha=0^{\rm o},~180^{\rm o}$). These features come from the polar cap regions (one from each hemisphere) which point periodically at the observer and result in a single peaked lightcurve. The other property of the low-energy radiation is that it exhibits a rather low degree of anisotropy; significant flux is seen at all viewing angles.

\begin{figure*}
\centering
\includegraphics[width=17cm]{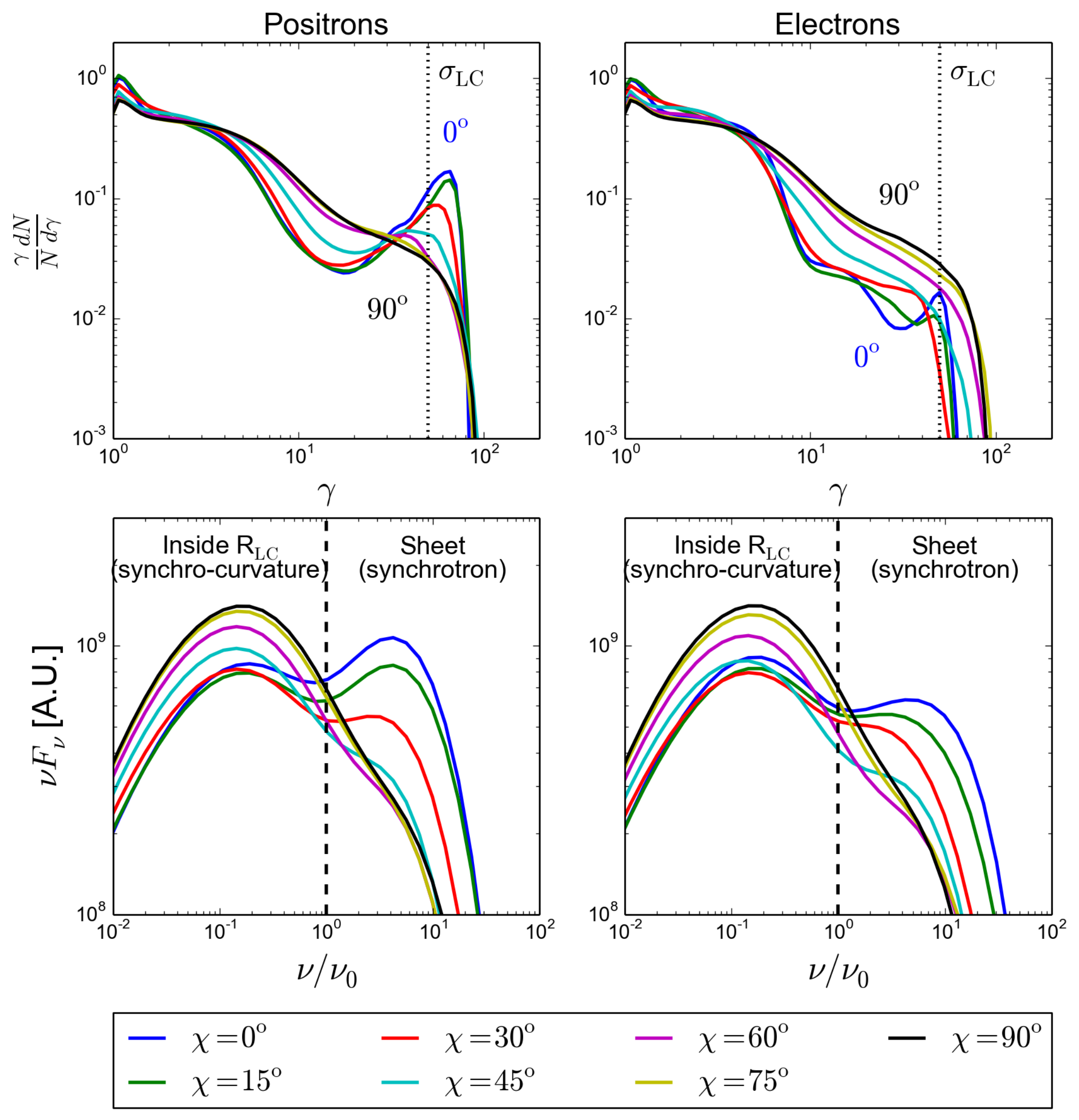}
\caption{Top panels: Total positron (left) and electron (right) spectra, $\left(\gamma/N \right)dN/d\gamma$, averaged over all directions as a function of the particle Lorentz factor $\gamma$ for the pulsar obliquities ranging from $\chi=0^{\rm o}$ to $90^{\rm o}$. The dotted line shows $\gamma=\sigma_{\rm LC}$. Bottom panels: Corresponding radiation spectral energy distributions ($\nu F_{\nu}$) emitted by positrons (left) and electrons (right), as a function of $\nu/\nu_0$. The vertical dashed line separates the low-energy ($\nu<\nu_0$) synchro-curvature radiation emitted within the light-cylinder radius, and the high-energy ($\nu>\nu_0$) synchrotron radiation from the equatorial current sheet.} 
\label{fig_spectra}
\end{figure*}

In contrast, the high-energy flux from the current sheet presents a dramatically different morphology. The skymap consists of a bright quasi-sinusoidal structure contained within $60^{\rm o}<\alpha< 120^{\rm o}$. These well-defined features can be observed only if strong radiative cooling is turned on in the simulation. An observer looking through the equator would then see a two-peaked lightcurve with some bridge emission, reminiscent of gamma-ray pulsars. Each peak happens when the current sheet sweeps across the observer's line of sight. To see this, we show in Figure~\ref{fig3d_light} where the observed high-energy emission comes from in the magnetosphere (the colored regions) at the phase of the first peak, $\Phi_{\rm P}=0.17$. In contrast to the total (isotropically averaged) emission (displayed in black and white, and in Figure~\ref{fig3d_flux}), the observed emission probes a small region of the current sheet. The positronic emission (top panel) forms a curved and narrow beam that extends from the light-cylinder radius all the way to the end of the box, with a maximum of emission within $1-2R_{\rm LC}$. In other words, this implies that different parts of the sheet radiate photons that arrive to the observer in phase to build the peak of the lightcurve, i.e., this is the signature of a caustic.

These results are robust against the pulsar obliquity. The left panels in Figure~\ref{fig_sky} show the high-energy skymaps for all the other inclinations simulated here. The sine-like structure is almost uniformly bright at low-inclinations ($\chi\lesssim 30^{\rm o}$), and breaks up into just two seemingly disconnected hot spots separated by $0.5$ in phase near the equator at high-inclinations ($\chi\gtrsim 45^{\rm o}$). A striking feature of the high-inclination solutions is that the emission is concentrated around the equatorial regions, even for the orthogonal rotator. This has important implications for the beaming factor of gamma-ray pulsars. High-inclination pulsars present another peculiarity, some of the equatorial wind region radiate energetic radiation beyond the light-cylinder, just before it reaches the current sheet. The wind leaves an imprint in the skymaps for $\chi\gtrsim 45^{\rm o}$ in the form of two extra hot spots at phases $\Phi_{\rm P}\approx 0.3$ and $0.8$ which results in extra secondary peaks in the lightcurves.

\begin{figure}
\centering
\includegraphics[width=8.6cm]{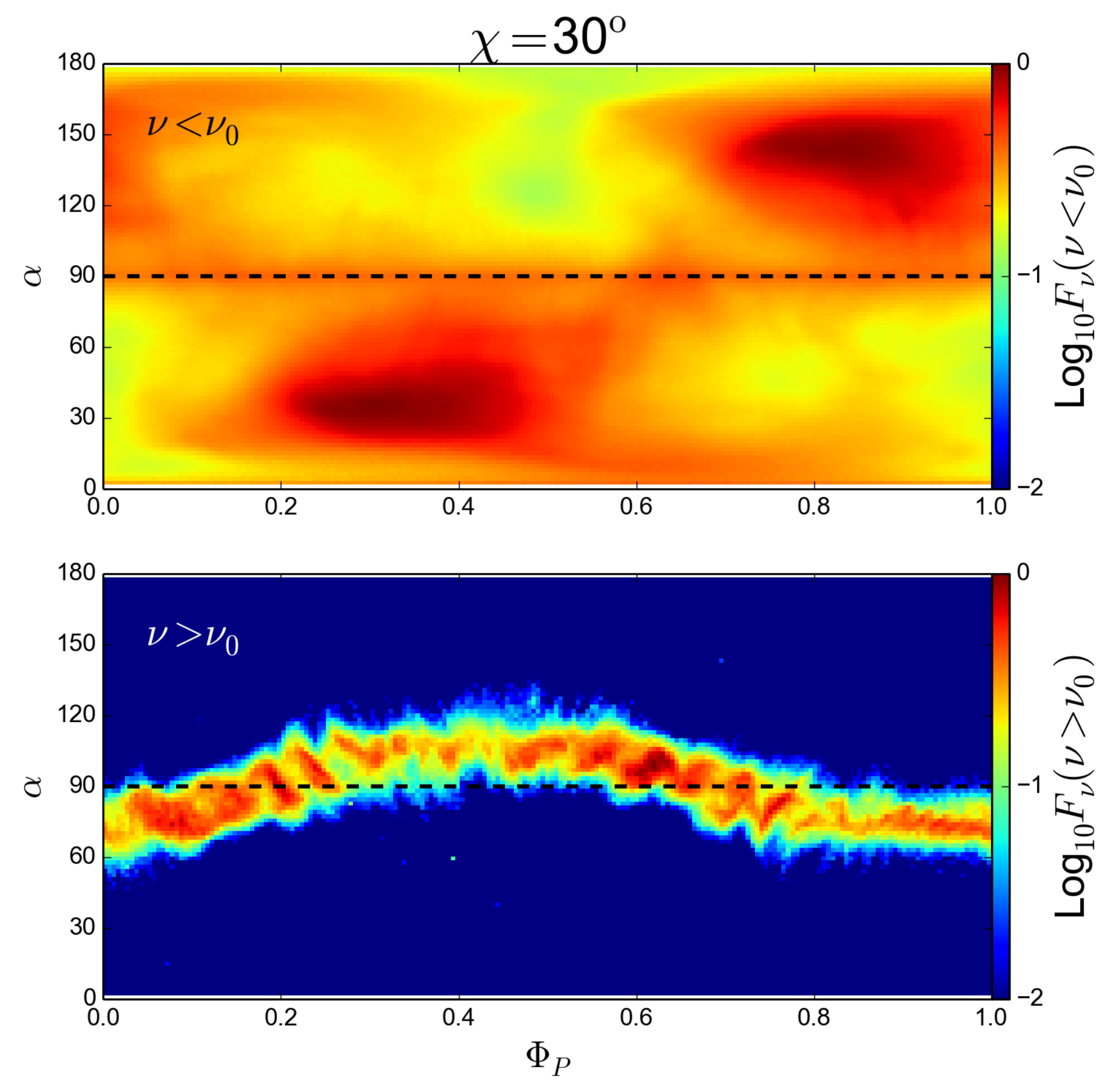}
\caption{Low- ($\nu<\nu_0$, top) and high-energy ($\nu>\nu_0$, bottom) radiation flux as function of the observer viewing angle $\alpha$ and the pulsar phase $\Phi_{\rm P}$, emitted by positrons for $\chi=30^{\rm o}$. Colors show the logarithm of the flux which is normalized to the maximum value of each skymap. The horizontal dashed black lines indicate an observer looking along the equator ($\alpha=90^{\rm o}$).} 
\label{fig_sky_freq}
\end{figure}

\begin{figure}
\centering
\includegraphics[width=8.6cm]{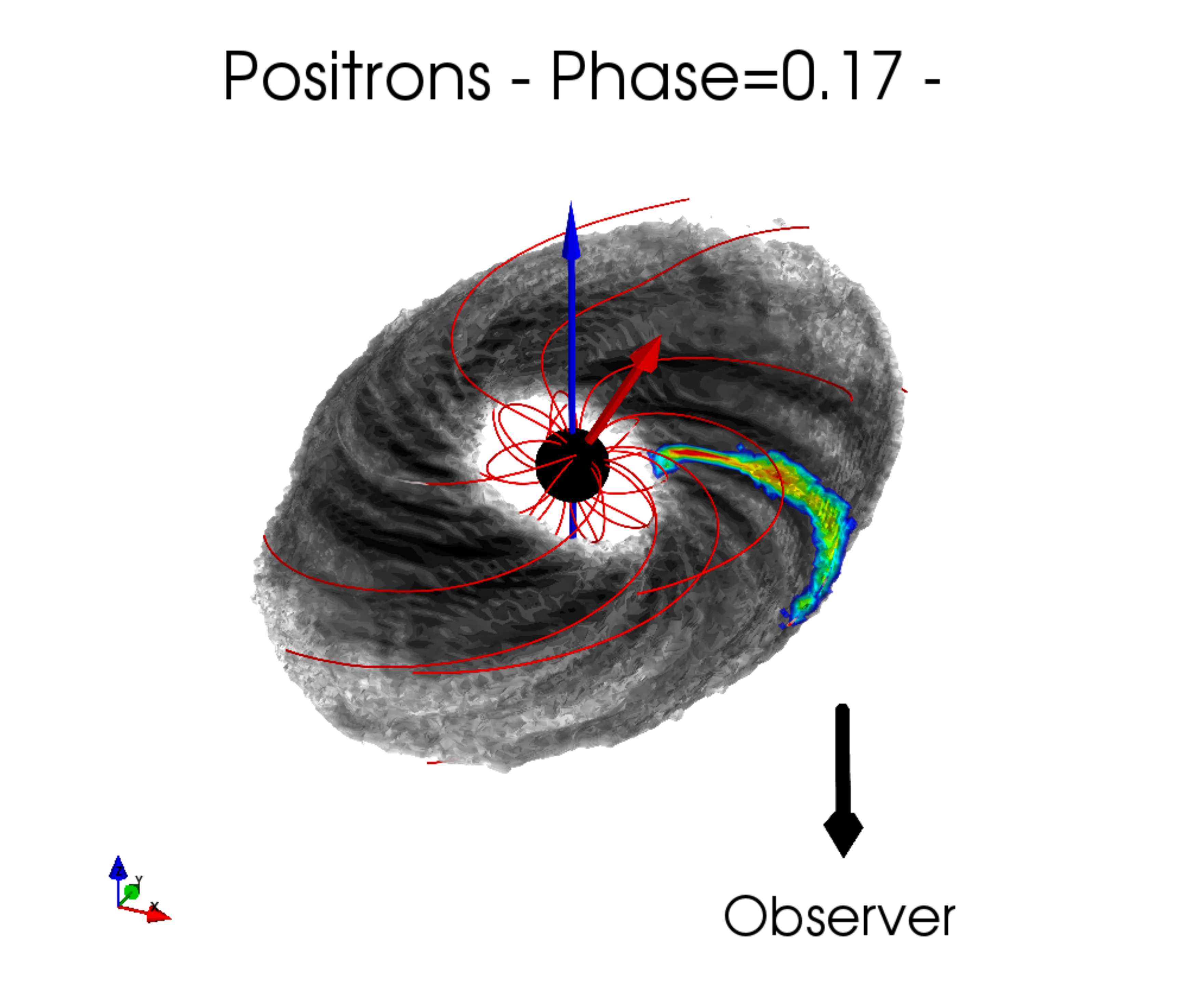}
\includegraphics[width=8.6cm]{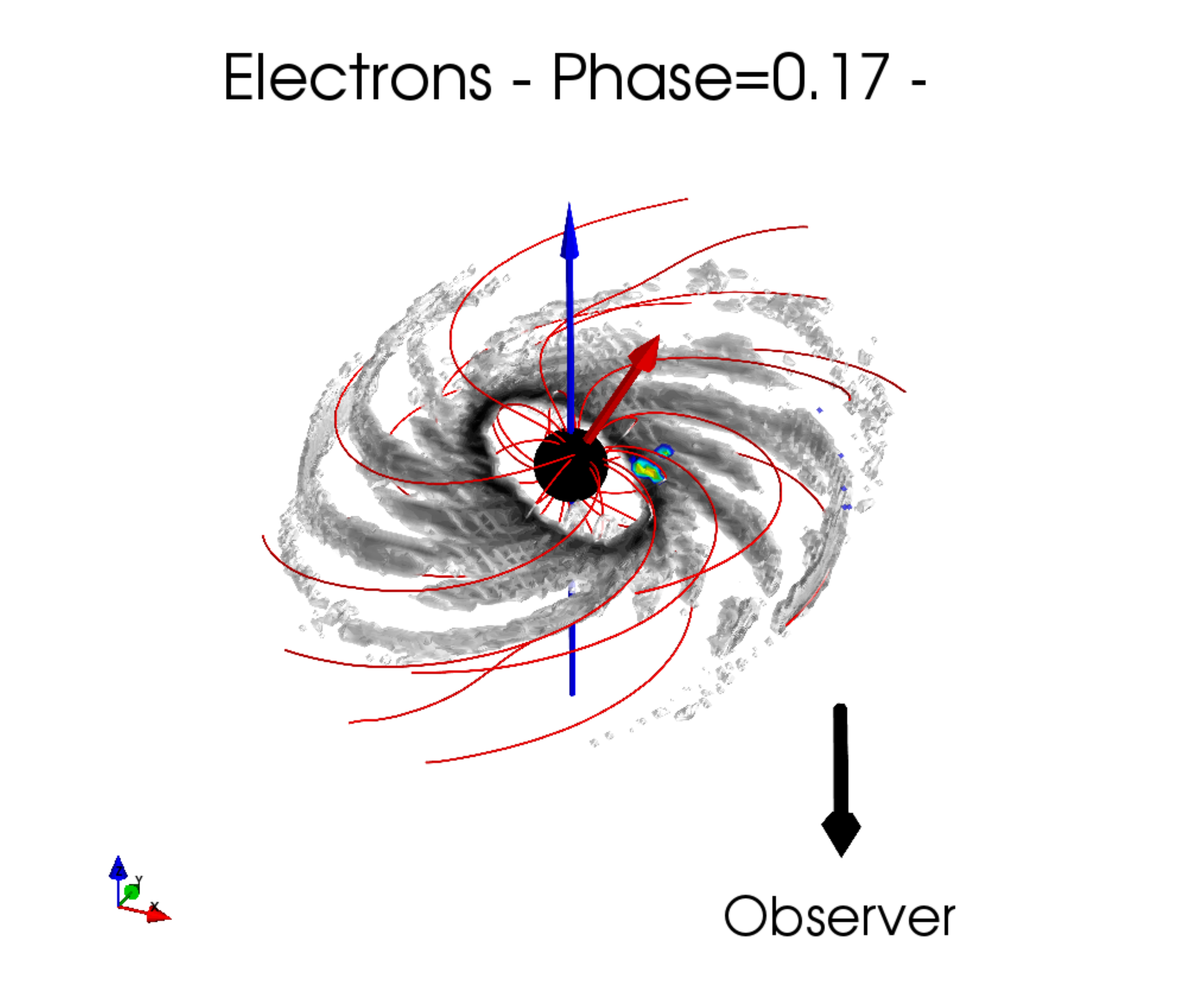}
\caption{In color: Three-dimensional rendering of the high-energy radiation flux from positrons (top) and electrons (bottom), as seen by a distant observer looking along the pulsar equator ($\alpha=90^{\rm o}$, direction indicated by the black arrow), at the pulsar phase $\Phi_{\rm P}=0.17$. The extended beam of positronic emission shows evidence of a caustic in the emission pattern from the current sheet. In back and white: The high-energy radiation emitted in all directions shown in Figure~\ref{fig3d_flux}. The red solid lines are the magnetic fields lines from Figure~\ref{fig3d_flux}. The arrows are along the magnetic (red) and the rotation (blue) axis of the pulsar (the black sphere) for $\chi=30^{\rm o}$. The radius varies from $r_{\star}$ to $3R_{\rm LC}$.} 
\label{fig3d_light}
\end{figure}

The above discussion applies only to energetic positrons, but electrons behave differently than positrons, and hence contribute to the observed flux differently. Indeed, as mentioned earlier and as shown in \citet{2015MNRAS.448..606C}, reconnection in the current sheet induces an electric field such that $\mathbf{E}\cdot\mathbf{B}\neq 0$, which pushes positrons outward and precipitates electrons towards the star through the Y-point and the separatrix current layers (see Figure~\ref{fig3d_orbits}). The counter-propagating electron beams result in additional structures in the high-energy radiation skymaps shown in the right panels of Figures~\ref{fig_sky}. At low inclinations, the skymaps contain bright emission along the equator which is the signature of electrons radiating at the Y-point at the base of the current sheet. This corresponds to the bright but compact emitting region shown in Figure~\ref{fig3d_light}, bottom panel. There are also filaments of emission connected to the Y-point, most noticeable at $\chi=15^{\rm o}$ and $30^{\rm o}$, pointing up to $\sim \pm 60^{\rm o}$ away from the equator. These features can also be explained by the electrons in the current sheet on their way back to the star. The largest deviations from the equator are coming from the most distant part of the current sheet (the spiral arms in Figure~\ref{fig3d_flux}, bottom panel). These deviations get smaller as the electrons get closer to the Y-point. For the special case of an orthogonal rotator, the electron skymap is the mirror image of the positron skymap with respect to the equator.

\begin{figure*}
\centering
\includegraphics[width=8.6cm]{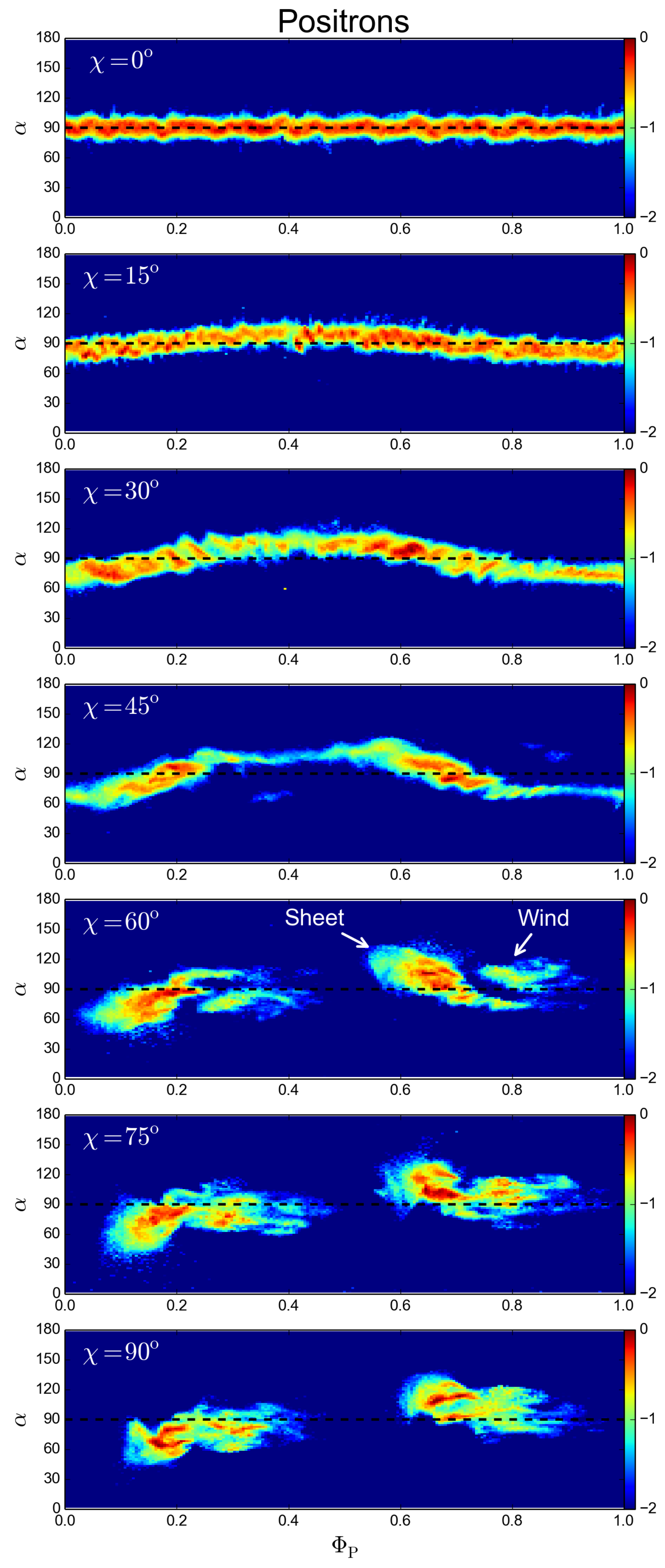}
\includegraphics[width=8.6cm]{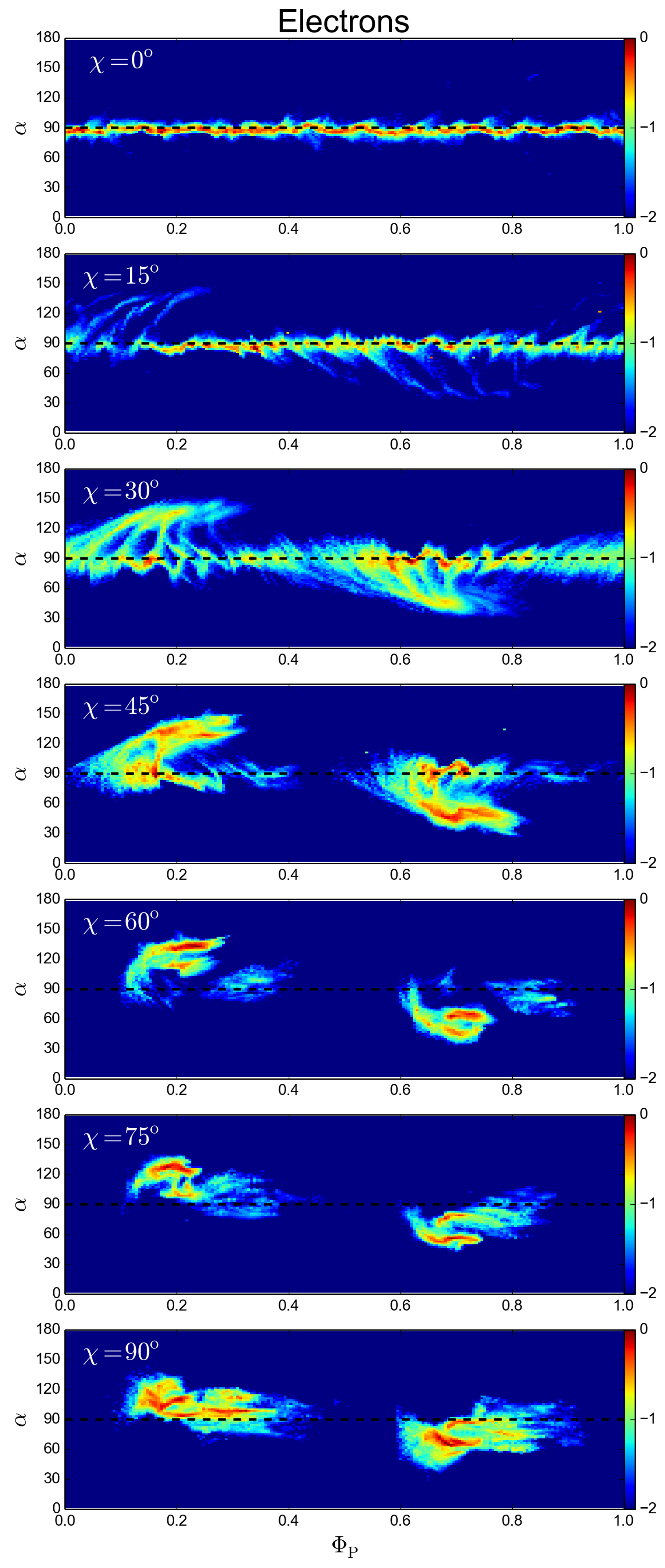}
\caption{High-energy skymaps ($\nu>\nu_0$) emitted by positrons (left) and electrons (right) for the following pulsar inclination angles (from top to bottom): $\chi=0^{\rm o}$, $15^{\rm o}$, $30^{\rm o}$, $45^{\rm o}$, $60^{\rm o}$, $75^{\rm o}$, $90^{\rm o}$. Colors show the ${\rm Log}_{10}$ of the integrated flux, normalized to its maximum value. The horizontal dashed black lines correspond to an observer looking along the equatorial plane ($\alpha=90^{\rm o}$). The arrows show the current sheet and wind contributions to the high-energy positron skymap for $\chi=60^{\rm o}$. These two components appear in all skymaps for $\chi\gtrsim 45^{\rm o}$.} 
\label{fig_sky}
\end{figure*}

\begin{figure*}
\centering
\includegraphics[width=17cm]{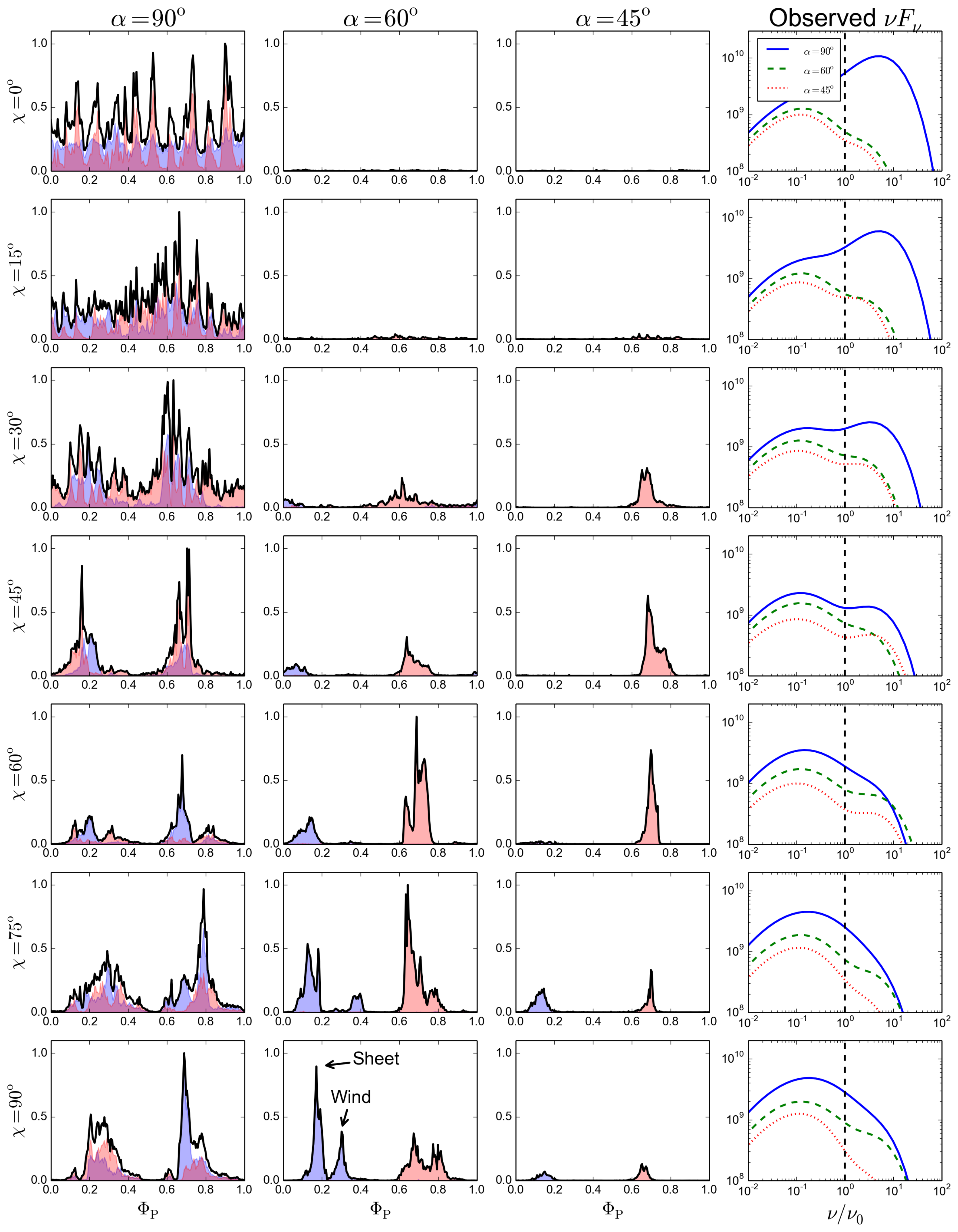}
\caption{First three columns: High-energy lightcurves for the viewing angles $\alpha=90^{\rm o}$ (left), $60^{\rm o}$ (middle) and $45^{\rm o}$ obtained from a 1D cut through the skymaps shown in Figure~\ref{fig_sky}, and for $\chi=0^{\rm o}$ (top), $15^{\rm o}$, $30^{\rm o}$, $45^{\rm o}$, $60^{\rm o}$, $75^{\rm o}$, $90^{\rm o}$ (bottom). The filled blue lines show the radiation flux emitted by positrons, the red filled lines show the emission from electrons, and the black solid lines give the sum from both species. In each panel, the total lightcurve is normalized to the maximum flux for a given inclination. The arrows show where the sheet and the wind contributions dominate for $\chi=90^{\rm o}$ and $\alpha=60^{\rm o}$. Last column: Phase-averaged radiation spectra ($\nu F_{\nu}$) emitted by both electrons and positrons observed at the viewing angles: $\alpha=90^{\rm o}$ (blue solid line), $60^{\rm o}$ (green dashed line), and $45^{\rm o}$ (red dotted line).}
\label{fig_lightcurves}
\end{figure*}

From the skymaps, one can generate any lightcurve by selecting a constant viewing angle $\alpha$ and varying $\Phi_{P}$. The first three columns in Figure~\ref{fig_lightcurves} present lightcurves for each inclination investigated here, for $\alpha=90^{\rm o}$, $60^{\rm o}$ and $45^{\rm o}$. The contribution from both species is shown with different colors, blue for the positronic emission and red for the electronic emission. The first immediate conclusion is that the two peak pattern is a very generic feature of the high-energy emission. It is present at all inclinations, with the exception of the aligned pulsar, by construction. In addition, the lightcurves exhibit substantial sub-pulse variability, visible as wiggles and irregularities in the skymaps (most notably at low inclinations). These small variations are due to plasma instabilities that grow in the current sheet (tearing and kink modes, see also Figure~3 in \citealt{2015ApJ...801L..19P}) which deform and fragment the layer into spiral structures (Figure~\ref{fig3d_flux}). However, it is likely that these fluctuations would disappear and leave two smoothed peaks if the lightcurves were averaged over multiple periods. For an observer looking along the equator ($\alpha=90^{\rm o}$), both electrons and positrons radiate approximatively in phase. The electronic emission is mostly coming from the Y-point, while the positronic emission comes from further away in the current sheet. At intermediate viewing angle ($\alpha=60^{\rm o}$), the first pulse is produced mainly by positrons and the second mainly by electrons. A single peak lightcurve can be obtained at lower viewing angle ($\alpha=45^{\rm o}$) and intermediate obliquities, where only electrons from the current sheet radiate. 

Because the high-energy radiation displays a pronounced anisotropy, the observed radiation spectrum can differ significantly from the total, isotropically-averaged spectra presented in Figure~\ref{fig_spectra}. The last column in Figure~\ref{fig_lightcurves} gives the phase-averaged spectra for the viewing angles, $\alpha=90^{\rm o}$, $60^{\rm o}$, and $45^{\rm o}$. The observed spectra present the largest changes with $\chi$ for a viewing angle $\alpha=90^{\rm o}$. At low inclinations, the spectra are dominated by the high-energy synchrotron component from the current sheet. In this case, the apparent isotropic high-energy radiative efficiency reaches up to $\eta^{\rm obs}_{\gamma}\approx 55\%$ due to beaming. The high-energy component decreases with increasing obliquity, and becomes smaller than the low-energy emission from the magnetosphere. At $\chi=30^{\rm o}-45^{\rm o}$, both components are comparable and give a characteristic two bumps aspect to the spectra.

\section{Discussion and conclusions}\label{sect_conclusion}

This study clearly indicates that the Y-point and the equatorial current sheet are the main sites for the emission of energetic radiation in the limit of a plasma-filled magnetosphere, a regime relevant to gamma-ray pulsars. Low-multiplicity solutions would also present gaps that would accelerate particles (e.g., \citealt{2015MNRAS.448..606C}), but those magnetospheres were not considered in this study.

Relativistic reconnection converts the magnetic energy into particle kinetic energy that is efficiently channeled into synchrotron radiation. This picture fits, at least qualitatively, the original radiating current sheet model by \citet{1996A&A...311..172L} and their extensions \citep{2002A&A...388L..29K, 2012MNRAS.424.2023P, 2013A&A...550A.101A, 2014ApJ...780....3U}, and also the dissipative force-free model of \citet{2014ApJ...793...97K} (their solution with ideal force-free inside $R_{\rm LC}$ and dissipative outside). The simulations show that most of particle acceleration and radiation occurs within a few light-cylinder radii, although we cannot exclude at this point that additional dissipation could occur further out, $r\gg R_{\rm LC}$, i.e., well outside the boundaries of our simulations. However, it is natural to expect that most of the high-energy radiation should originate not too far from the light cylinder where the magnetic field is strongest. In addition, because of the formation of caustics in the current sheet, extending the size of the emitting zone would increase excessively the amplitude of the gamma-ray peak in the lightcurve. 

In most cases, the simulated high-energy lightcurves present the two peaks structure reminiscent of observed gamma-ray pulsars. A single peak can also be observed if looking at higher latitude. The peaks in the lightcurve occur when our line of sight passes through the current sheet. For an oblique pulsar, this happens twice per period if looking along the equator of the pulsar. Hence, the gamma-ray lightcurve is shaped by the geometry of the current sheet only. On top of these main peaks of geometric origin, we found sub-pulse variability which is provoked by plasma waves and instabilities that deform and fragment the current sheet in the form of kinked spiral structures, most visible at low obliquity ($\chi\lesssim 45^{\rm o}$, see also \citealt{2015ApJ...801L..19P}). These variations in plasma density, and hence, in high-energy radiation leave an imprint in the lightcurves which look like small and rapid flickering. These plasma instabilities could be similar to the tearing and kink modes usually observed in local 3D PIC simulations of relativistic reconnection (e.g., \citealt{2008ApJ...677..530Z, 2014ApJ...782..104C, 2014ApJ...783L..21S}). If so, they should form on the plasma scale, of order the collisionless skindepth and Larmor radius of the particles which are microscopic quantities in real pulsars. As a result, it is most likely that these fluctuations are exaggerated in the simulations where the separation of scale is limited, and would not be observable in practice. In addition, measured gamma-ray lightcurves are typically reconstructed after a large number of rotation periods due to flux limitations \citep{2010ApJS..187..460A}, which would completely average out any random fluctuations. We will construct such averages using longer simulations in the future.

We found that high-obliquity ($\chi\gtrsim 45^{\rm o}$) solutions contains significant energetic radiation coming from the wind region that feeds the current layer with plasma. This acceleration could be induced by reconnection in the current sheet via the $\mathbf{E}\times\mathbf{B}$ drift, where $\mathbf{E}$ is the reconnection electric field. While this acceleration seems robust and in agreement with previous MHD simulations \citep{2013MNRAS.435L...1T}, it is not clear at this point whether this emission should be observable in the same energy band as the synchrotron emission from the current sheet. The separation in energy scale between the particles in the sheet (given by $\sigma_{\rm LC}$) and the particles accelerated in the wind (related to $\mathbf{E}\times\mathbf{B}$ drift) is too small in our simulations due to limited computational power, to clearly disentangle both contributions. In the simulated lightcurves, the emission coming from the wind produces one secondary peak following the main peak, so that in principle the lightcurve could have up to 4 peaks per rotation period. The simulated skymaps show another important feature. Regardless of the pulsar obliquity, the high-energy radiation is always concentrated around the equatorial regions, even for the orthogonal rotator, because this is where most of the power released by the pulsar is concentrated and dissipated (see e.g., the latest studies by \citealt{2013MNRAS.435L...1T, 2015arXiv150301467T}).

The two particle species have two different radiating patterns, because of their different acceleration history in the current sheet. Both species are accelerated by the reconnection electric field in the current sheet, which makes positrons accelerate outward while energetic electrons precipitate back to the star for $\boldsymbol{\Omega}\cdot\mathbf{B}>0$ on the pole \citep{2015MNRAS.448..606C}. Therefore, the positronic emission is spread over the whole current layer, within $1-2 R_{\rm LC}$, while the electronic emission is mostly concentrated at the Y-point, where both the particle energy and the magnetic field strength are highest. These differences appear in the particle and radiation spectra of both species. The particle and radiation spectra exhibit large variations that depends on the viewing angle and pulsar obliquity. In most cases, the radiation spectrum is composed of two bumps, a low-energy bump of synchro-curvature radiation emitted within the light-cylinder, and a high-energy synchrotron bump from the current sheet. The high-energy component peaks around $10\nu_0$, which corresponds to particles with $\gamma=\sigma_{\rm LC}$ immersed in a perpendicular magnetic field $\tilde{B}_{\perp}\approx 0.1 B_{\rm LC}$, where $B_{\rm LC}\approx B_{\star}\left(r_{\star}/R_{\rm LC}\right)^3$ is the magnetic field at the light cylinder, i.e., $\nu_{\rm max}\approx 3 e \left(0.1 B_{\rm LC}\right)\sigma^2_{\rm LC}/4\pi m_{\rm e} c$. Applying this relation to a $B_{\star}=10^{12}~$G magnetic field, and $\sigma_{\rm LC}\sim 10^6$ for a typical energetic young pulsar yields $\nu_{\rm cut}\approx 4$GeV, which is a very common break energy in gamma-ray pulsars. However, more simulations are needed to uncover the exact dependence of the cutoff energy with the magnetic field strength at the light-cylinder. In addition, these simulations should include the full pair formation physics which may have an effect on the shape of the particle and radiation spectra. Simulations with pair formation indicate that significant particle acceleration also occurs within the return current layers \citep{2014ApJ...795L..22C, 2015ApJ...801L..19P}. Hence, it is not excluded that additional high-energy radiation could come from within the light-cylinder.

The high-energy emission represents a significant fraction of the pulsar spindown power, varying from a few percent for highly inclined pulsars ($\chi\gtrsim 60^{\rm o}$) to almost $10\%$ for the aligned pulsar. Hence, low-inclination pulsars are better at accelerating particles and at emitting energetic radiation than highly-inclined pulsars. The expected radiative efficiencies are compatible with gamma-ray pulsars seen by the \emph{Fermi}-LAT \citep{2010ApJS..187..460A}. However, note that higher/lower radiative efficiencies can be observed if looking within/outside the pulsar equatorial regions due to the strong beaming of the high-energy emission. This study was not designed to apply our results to actual gamma-ray observations, but instead this should be seen as a first step towards this objective, that demonstrates the feasibility and the suitability of the fully kinetic approach to this problem. Another study should be specially dedicated to the comparison of the PIC simulations directly with observations, e.g., including spectral and lightcurve fitting and polarization.

\section*{Acknowledgments}

BC acknowledges support from the Lyman Spitzer Jr. Fellowship awarded by the Department of Astrophysical Sciences at Princeton University. This work was supported by the Max-Planck/Princeton Center for Plasma Physics, by NASA Earth and Space Science Fellowship Program (grant NNX15AT50H to AP), by NASA grants NNX14AQ67G and NNX15AM30G, and by the Simons Foundation (grant 291817 to AS). Computing resources were provided by the PICSciE-OIT TIGRESS High Performance Computing Center and Visualization Laboratory at Princeton University, and by the TACC on Stampede via the XSEDE allocations TG-PHY140041 and AST140037.

%%%%%%%%%%%%%%%%%%%%%%%%%%%%%%%%%%%%%%%%%%%%%%%%%%



\bibliographystyle{aa}
\bibliography{pulsar_light} % if your bibtex file is called example.bib

\begin{thebibliography}{51}
\expandafter\ifx\csname natexlab\endcsname\relax\def\natexlab#1{#1}\fi

\bibitem[{{Abdo} {et~al.}(2010){Abdo}, {Ackermann}, {Ajello}, {Atwood},
  {Axelsson}, {Baldini}, {Ballet}, {Barbiellini}, {Baring}, {Bastieri}, \&
  et~al.}]{2010ApJS..187..460A}
{Abdo}, A.~A., {Ackermann}, M., {Ajello}, M., {et~al.} 2010, \apjs, 187, 460

\bibitem[{{Abdo} {et~al.}(2013){Abdo}, {Ajello}, {Allafort}, {Baldini},
  {Ballet}, {Barbiellini}, {Baring}, {Bastieri}, {Belfiore}, {Bellazzini}, \&
  et~al.}]{2013ApJS..208...17A}
{Abdo}, A.~A., {Ajello}, M., {Allafort}, A., {et~al.} 2013, \apjs, 208, 17

\bibitem[{{Arka} \& {Dubus}(2013)}]{2013A&A...550A.101A}
{Arka}, I. \& {Dubus}, G. 2013, \aap, 550, A101

\bibitem[{{Arons}(1983)}]{1983ApJ...266..215A}
{Arons}, J. 1983, \apj, 266, 215

\bibitem[{{Arons} \& {Scharlemann}(1979)}]{1979ApJ...231..854A}
{Arons}, J. \& {Scharlemann}, E.~T. 1979, \apj, 231, 854

\bibitem[{{Bai} \& {Spitkovsky}(2010{\natexlab{a}})}]{2010ApJ...715.1282B}
{Bai}, X.-N. \& {Spitkovsky}, A. 2010{\natexlab{a}}, \apj, 715, 1282

\bibitem[{{Bai} \& {Spitkovsky}(2010{\natexlab{b}})}]{2010ApJ...715.1270B}
{Bai}, X.-N. \& {Spitkovsky}, A. 2010{\natexlab{b}}, \apj, 715, 1270

\bibitem[{{Belyaev}(2015)}]{2015MNRAS.449.2759B}
{Belyaev}, M.~A. 2015, \mnras, 449, 2759

\bibitem[{{Blumenthal} \& {Gould}(1970)}]{1970RvMP...42..237B}
{Blumenthal}, G.~R. \& {Gould}, R.~J. 1970, Reviews of Modern Physics, 42, 237

\bibitem[{{Cerutti} {et~al.}(2015){Cerutti}, {Philippov}, {Parfrey}, \&
  {Spitkovsky}}]{2015MNRAS.448..606C}
{Cerutti}, B., {Philippov}, A., {Parfrey}, K., \& {Spitkovsky}, A. 2015,
  \mnras, 448, 606

\bibitem[{{Cerutti} {et~al.}(2013){Cerutti}, {Werner}, {Uzdensky}, \&
  {Begelman}}]{2013ApJ...770..147C}
{Cerutti}, B., {Werner}, G.~R., {Uzdensky}, D.~A., \& {Begelman}, M.~C. 2013,
  \apj, 770, 147

\bibitem[{{Cerutti} {et~al.}(2014){Cerutti}, {Werner}, {Uzdensky}, \&
  {Begelman}}]{2014ApJ...782..104C}
{Cerutti}, B., {Werner}, G.~R., {Uzdensky}, D.~A., \& {Begelman}, M.~C. 2014,
  \apj, 782, 104

\bibitem[{{Chen} \& {Beloborodov}(2014)}]{2014ApJ...795L..22C}
{Chen}, A.~Y. \& {Beloborodov}, A.~M. 2014, \apjl, 795, L22

\bibitem[{{Cheng} {et~al.}(1986{\natexlab{a}}){Cheng}, {Ho}, \&
  {Ruderman}}]{1986ApJ...300..500C}
{Cheng}, K.~S., {Ho}, C., \& {Ruderman}, M. 1986{\natexlab{a}}, \apj, 300, 500

\bibitem[{{Cheng} {et~al.}(1986{\natexlab{b}}){Cheng}, {Ho}, \&
  {Ruderman}}]{1986ApJ...300..522C}
{Cheng}, K.~S., {Ho}, C., \& {Ruderman}, M. 1986{\natexlab{b}}, \apj, 300, 522

\bibitem[{{Cheng} \& {Zhang}(1996)}]{1996ApJ...463..271C}
{Cheng}, K.~S. \& {Zhang}, J.~L. 1996, \apj, 463, 271

\bibitem[{{Contopoulos}(2007)}]{2007A&A...472..219C}
{Contopoulos}, I. 2007, \aap, 472, 219

\bibitem[{{Coroniti}(1990)}]{1990ApJ...349..538C}
{Coroniti}, F.~V. 1990, \apj, 349, 538

\bibitem[{{Daugherty} \& {Harding}(1982)}]{1982ApJ...252..337D}
{Daugherty}, J.~K. \& {Harding}, A.~K. 1982, \apj, 252, 337

\bibitem[{{Dyks} \& {Rudak}(2003)}]{2003ApJ...598.1201D}
{Dyks}, J. \& {Rudak}, B. 2003, \apj, 598, 1201

\bibitem[{{Goldreich} \& {Julian}(1969)}]{1969ApJ...157..869G}
{Goldreich}, P. \& {Julian}, W.~H. 1969, \apj, 157, 869

\bibitem[{{Harding} {et~al.}(1978){Harding}, {Tademaru}, \&
  {Esposito}}]{1978ApJ...225..226H}
{Harding}, A.~K., {Tademaru}, E., \& {Esposito}, L.~W. 1978, \apj, 225, 226

\bibitem[{{Holland}(1983)}]{1983ITNS...30.4592H}
{Holland}, R. 1983, IEEE Transactions on Nuclear Science, 30, 4592

\bibitem[{{Kalapotharakos} {et~al.}(2014){Kalapotharakos}, {Harding}, \&
  {Kazanas}}]{2014ApJ...793...97K}
{Kalapotharakos}, C., {Harding}, A.~K., \& {Kazanas}, D. 2014, \apj, 793, 97

\bibitem[{{Kalapotharakos} {et~al.}(2012){Kalapotharakos}, {Harding},
  {Kazanas}, \& {Contopoulos}}]{2012ApJ...754L...1K}
{Kalapotharakos}, C., {Harding}, A.~K., {Kazanas}, D., \& {Contopoulos}, I.
  2012, \apjl, 754, L1

\bibitem[{{Kelner} {et~al.}(2015){Kelner}, {Prosekin}, \&
  {Aharonian}}]{2015AJ....149...33K}
{Kelner}, S.~R., {Prosekin}, A.~Y., \& {Aharonian}, F.~A. 2015, \aj, 149, 33

\bibitem[{{Kirk}(2004)}]{2004PhRvL..92r1101K}
{Kirk}, J.~G. 2004, Physical Review Letters, 92, 181101

\bibitem[{{Kirk} {et~al.}(2002){Kirk}, {Skj{\ae}raasen}, \&
  {Gallant}}]{2002A&A...388L..29K}
{Kirk}, J.~G., {Skj{\ae}raasen}, O., \& {Gallant}, Y.~A. 2002, \aap, 388, L29

\bibitem[{{Li} {et~al.}(2012){Li}, {Spitkovsky}, \&
  {Tchekhovskoy}}]{2012ApJ...746...60L}
{Li}, J., {Spitkovsky}, A., \& {Tchekhovskoy}, A. 2012, \apj, 746, 60

\bibitem[{{Lyubarskii}(1996)}]{1996A&A...311..172L}
{Lyubarskii}, Y.~E. 1996, \aap, 311, 172

\bibitem[{{Mochol} \& {P{\'e}tri}(2015)}]{2015MNRAS.449L..51M}
{Mochol}, I. \& {P{\'e}tri}, J. 2015, \mnras, 449, L51

\bibitem[{{Muslimov} \& {Harding}(2003)}]{2003ApJ...588..430M}
{Muslimov}, A.~G. \& {Harding}, A.~K. 2003, \apj, 588, 430

\bibitem[{{Muslimov} \& {Harding}(2004)}]{2004ApJ...606.1143M}
{Muslimov}, A.~G. \& {Harding}, A.~K. 2004, \apj, 606, 1143

\bibitem[{{P{\'e}tri}(2012)}]{2012MNRAS.424.2023P}
{P{\'e}tri}, J. 2012, \mnras, 424, 2023

\bibitem[{{Philippov} \& {Spitkovsky}(2014)}]{2014ApJ...785L..33P}
{Philippov}, A.~A. \& {Spitkovsky}, A. 2014, \apjl, 785, L33

\bibitem[{{Philippov} {et~al.}(2015){Philippov}, {Spitkovsky}, \&
  {Cerutti}}]{2015ApJ...801L..19P}
{Philippov}, A.~A., {Spitkovsky}, A., \& {Cerutti}, B. 2015, \apjl, 801, L19

\bibitem[{{Prosekin} {et~al.}(2013){Prosekin}, {Kelner}, \&
  {Aharonian}}]{2013arXiv1305.0783P}
{Prosekin}, A.~Y., {Kelner}, S.~R., \& {Aharonian}, F.~A. 2013, ArXiv e-prints

\bibitem[{{Romani} \& {Yadigaroglu}(1995)}]{1995ApJ...438..314R}
{Romani}, R.~W. \& {Yadigaroglu}, I.-A. 1995, \apj, 438, 314

\bibitem[{{Ruderman} \& {Sutherland}(1975)}]{1975ApJ...196...51R}
{Ruderman}, M.~A. \& {Sutherland}, P.~G. 1975, \apj, 196, 51

\bibitem[{{Sironi} \& {Spitkovsky}(2014)}]{2014ApJ...783L..21S}
{Sironi}, L. \& {Spitkovsky}, A. 2014, \apjl, 783, L21

\bibitem[{{Spitkovsky}(2006)}]{2006ApJ...648L..51S}
{Spitkovsky}, A. 2006, \apjl, 648, L51

\bibitem[{{Sturrock}(1971)}]{1971ApJ...164..529S}
{Sturrock}, P.~A. 1971, \apj, 164, 529

\bibitem[{{Tamburini} {et~al.}(2010){Tamburini}, {Pegoraro}, {Di Piazza},
  {Keitel}, \& {Macchi}}]{2010NJPh...12l3005T}
{Tamburini}, M., {Pegoraro}, F., {Di Piazza}, A., {Keitel}, C.~H., \& {Macchi},
  A. 2010, New Journal of Physics, 12, 123005

\bibitem[{{Tchekhovskoy} {et~al.}(2015){Tchekhovskoy}, {Philippov}, \&
  {Spitkovsky}}]{2015arXiv150301467T}
{Tchekhovskoy}, A., {Philippov}, A., \& {Spitkovsky}, A. 2015, ArXiv e-prints

\bibitem[{{Tchekhovskoy} {et~al.}(2013){Tchekhovskoy}, {Spitkovsky}, \&
  {Li}}]{2013MNRAS.435L...1T}
{Tchekhovskoy}, A., {Spitkovsky}, A., \& {Li}, J.~G. 2013, \mnras, 435, L1

\bibitem[{{Timokhin} \& {Arons}(2013)}]{2013MNRAS.429...20T}
{Timokhin}, A.~N. \& {Arons}, J. 2013, \mnras, 429, 20

\bibitem[{{Uzdensky}(2003)}]{2003ApJ...598..446U}
{Uzdensky}, D.~A. 2003, \apj, 598, 446

\bibitem[{{Uzdensky} \& {McKinney}(2011)}]{2011PhPl...18d2105U}
{Uzdensky}, D.~A. \& {McKinney}, J.~C. 2011, Physics of Plasmas, 18, 042105

\bibitem[{{Uzdensky} \& {Spitkovsky}(2014)}]{2014ApJ...780....3U}
{Uzdensky}, D.~A. \& {Spitkovsky}, A. 2014, \apj, 780, 3

\bibitem[{{Vigan{\`o}} {et~al.}(2015){Vigan{\`o}}, {Torres}, {Hirotani}, \&
  {Pessah}}]{2015MNRAS.447.1164V}
{Vigan{\`o}}, D., {Torres}, D.~F., {Hirotani}, K., \& {Pessah}, M.~E. 2015,
  \mnras, 447, 1164

\bibitem[{{Zenitani} \& {Hoshino}(2008)}]{2008ApJ...677..530Z}
{Zenitani}, S. \& {Hoshino}, M. 2008, \apj, 677, 530

\end{thebibliography}

\appendix

\section{Cell-integrated expression of Maxwell's equations on the 3D spherical Yee grid}\label{app_maxwell}

\begin{figure}
\centering
\includegraphics[width=8cm]{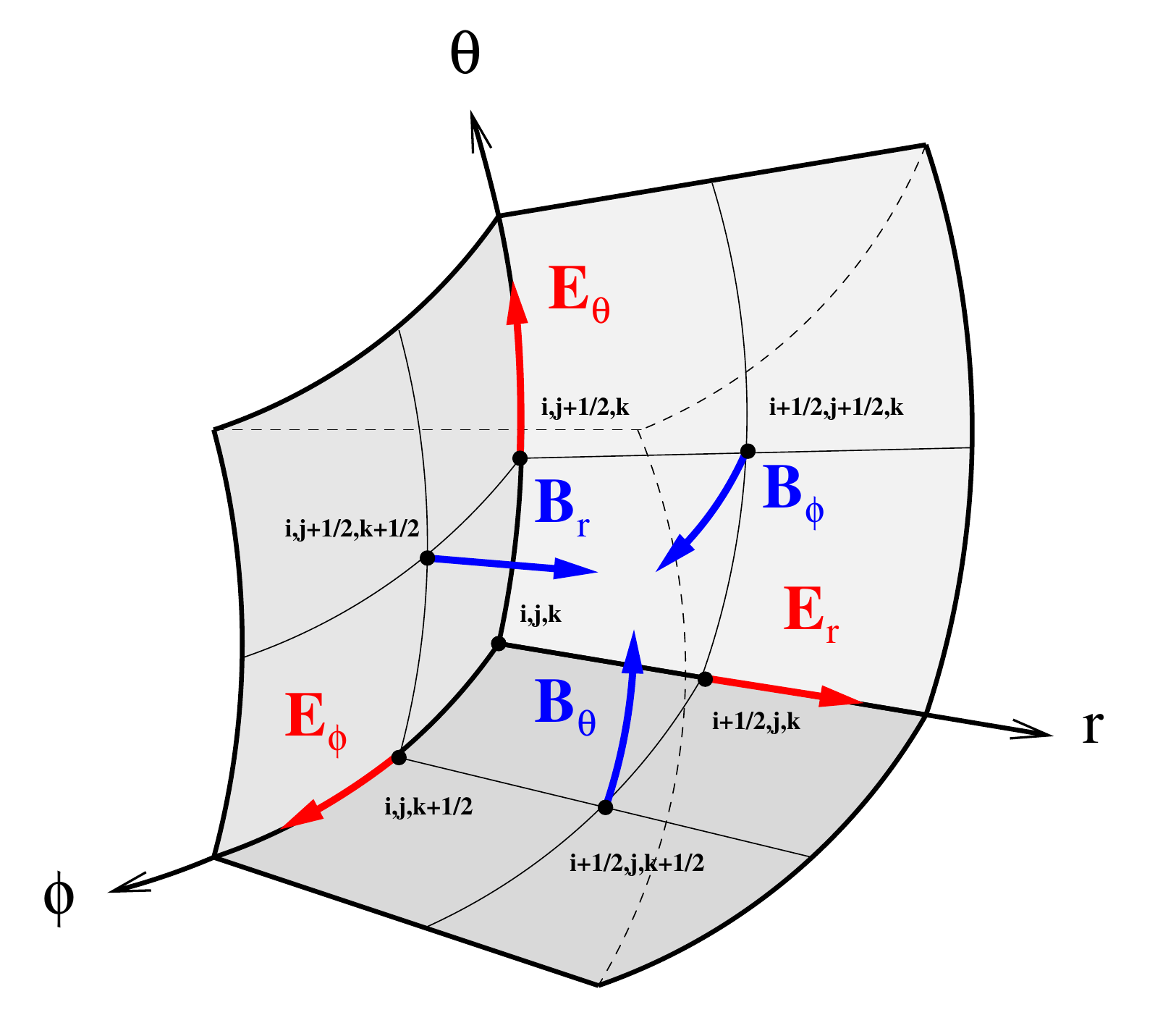}
\caption{Three-dimensional spherical Yee-lattice used in {\tt ZELTRON}.}
\label{fig_3dyee}
\end{figure}

Here, we present the full 3D integrated expressions for $\mathbf{\nabla}\times \mathbf{E}$, and $\nabla\times \mathbf{B}$ as used in {\tt ZELTRON}, where the field components are defined on the standard staggered Yee-lattice (see Figure~\ref{fig_3dyee}). The cells are labeled by the triplet of integers (i,j,k), for the $r$-,~$\theta$-, and $\phi$-direction respectively. Using Stokes' theorem on the faces of the cell (i,j,k) yields
\begin{equation}
\iint_{\mathcal{S}}(\mathbf{\nabla}\times\mathbf{E})\cdot d{\bf \mathcal{S}}=\oint_{\mathcal C}\mathbf{E}\cdot d\bf{\mathcal{C}},
\label{stokes}
\end{equation}
where $\bf{\mathcal{S}}$ is the surface vector pointing away from the cell, and $\bf{\mathcal{C}}$ is the contour vector circulating around the cell. On the Yee-lattice, the components of $\mathbf{\nabla}\times \mathbf{E}$ and $\mathbf{\nabla}\times \mathbf{B}$ are given by
\begin{eqnarray}
\left(\mathbf{\nabla}\times\mathbf{E}\right)_{\rm r_{\rm i,j+\frac{1}{2},k+\frac{1}{2}}} &=& \frac{\left(\sin\theta_{\rm j+1}E_{\phi_{\rm i,j+1,k+\frac{1}{2}}}-\sin\theta_{\rm j}E_{\phi_{\rm i,j,k+\frac{1}{2}}}\right)}{r_{\rm i}\Delta\mu_{\rm j+\frac{1}{2}}} \nonumber \\
&-& \frac{\Delta\theta\left(E_{\theta_{\rm i,j+\frac{1}{2},k+1}}-E_{\theta_{\rm i,j+\frac{1}{2},k}}\right)}{r_{\rm i}\Delta\mu_{\rm j+\frac{1}{2}}\Delta\phi}
\end{eqnarray}
\begin{eqnarray}
\left(\mathbf{\nabla}\times\mathbf{E}\right)_{\rm{\theta_{i+\frac{1}{2},j,k+\frac{1}{2}}}} &=& -\frac{2\left(r_{\rm i+1}E_{\phi_{\rm i+1,j,k+\frac{1}{2}}}-r_{\rm i}E_{\phi_{\rm i,j,k+\frac{1}{2}}}\right)}{\Delta r^2_{\rm i+\frac{1}{2}}} \nonumber \\
&+&\frac{2\Delta r_{\rm i+\frac{1}{2}}\left(E_{\rm r_{\rm i+\frac{1}{2},j,k+1}}-E_{\rm r_{\rm i+\frac{1}{2},j,k}}\right)}{\Delta r^2_{\rm i+\frac{1}{2}}\sin\theta_{\rm j}\Delta\phi}
\end{eqnarray}
\begin{eqnarray}
\left(\mathbf{\nabla}\times\mathbf{E}\right)_{\phi_{\rm i+\frac{1}{2},j+\frac{1}{2},k}} &=& \frac{2\left(r_{\rm i+1}E_{\theta_{\rm i+1,j+\frac{1}{2},k}}-r_{\rm i}E_{\theta_{\rm i,j+\frac{1}{2},k}}\right)}{\Delta r^2_{\rm i+\frac{1}{2}}} \nonumber \\
&-& \frac{2\Delta r_{\rm i+\frac{1}{2}}\left(E_{\rm r_{\rm i+\frac{1}{2},j+1,k}}-
E_{\rm r_{\rm i+\frac{1}{2},j,k}}\right)}{\Delta r^2_{\rm i+\frac{1}{2}}\Delta\theta}
\end{eqnarray}
\begin{eqnarray}
\left(\mathbf{\nabla}\times\mathbf{B}\right)_{\rm r_{\rm i+\frac{1}{2},j,k}} &=& \frac{\left(\sin\theta_{\rm j+\frac{1}{2}}B_{\phi_{\rm i+\frac{1}{2},j+\frac{1}{2},k}}-\sin\theta_{\rm j-\frac{1}{2}}B_{\phi_{\rm i+\frac{1}{2},j-\frac{1}{2},k}}\right)}{r_{\rm i+\frac{1}{2}}\Delta\mu_{\rm j}} \nonumber \\
&-& \frac{\Delta\theta\left(B_{\theta_{\rm i+\frac{1}{2},j,k+\frac{1}{2}}}-B_{\theta_{\rm i+\frac{1}{2},j,k-\frac{1}{2}}}\right)}{r_{\rm i+\frac{1}{2}}\Delta\mu_{\rm j}\Delta\phi}
\end{eqnarray}
\begin{eqnarray}
\left(\mathbf{\nabla}\times\mathbf{B}\right)_{\rm{\theta_{i,j+\frac{1}{2},k}}} &=& -\frac{2\left(r_{\rm i+\frac{1}{2}}B_{\phi_{\rm i+\frac{1}{2},j+\frac{1}{2},k}}-r_{\rm i-\frac{1}{2}}B_{\phi_{\rm i-\frac{1}{2},j+\frac{1}{2},k}}\right)}{\Delta r^2_{\rm i}} \nonumber \\
&+& \frac{2\Delta r_{\rm i}\left(B_{\rm r_{\rm i,j+\frac{1}{2},k+\frac{1}{2}}}-B_{\rm r_{\rm i,j+\frac{1}{2},k-\frac{1}{2}}}\right)}{\Delta r^2_{\rm i}\sin\theta_{\rm j+\frac{1}{2}}\Delta\phi}
\end{eqnarray}
\begin{eqnarray}
\left(\mathbf{\nabla}\times\mathbf{B}\right)_{\phi_{\rm i,j,k+\frac{1}{2}}} &=& \frac{2\left(r_{\rm i+\frac{1}{2}}B_{\theta_{\rm i+\frac{1}{2},j,k+\frac{1}{2}}}-r_{\rm i-\frac{1}{2}}B_{\theta_{\rm i-\frac{1}{2},j,k+\frac{1}{2}}}\right)}{\Delta r^2_{\rm i}} \nonumber \\
&-& \frac{2\Delta r_{\rm i}\left(B_{\rm r_{\rm i,j+\frac{1}{2},k+\frac{1}{2}}}-
B_{\rm r_{\rm i,j-\frac{1}{2},k+\frac{1}{2}}}\right)}{\Delta r^2_{\rm i}\Delta\theta},
\end{eqnarray}
where,
\begin{eqnarray}
\Delta r_{\rm i} &=& \left(r_{\rm i+\frac{1}{2}}-r_{\rm i-\frac{1}{2}}\right)\\
\Delta r^2_{\rm i} &=& \left(r^2_{\rm i+\frac{1}{2}}-r^2_{\rm i-\frac{1}{2}}\right)\\
\Delta r_{\rm i+\frac{1}{2}} &=& \left(r_{\rm i+1}-r_{\rm i}\right)\\
\Delta r^2_{\rm i+\frac{1}{2}} &=& \left(r^2_{\rm i+1}-r^2_{\rm i}\right)\\
\Delta\mu_{\rm j} &=& \left(\cos\theta_{\rm j-\frac{1}{2}}-\cos\theta_{\rm j+\frac{1}{2}}\right)\\
\Delta\mu_{\rm j+\frac{1}{2}} &=& \left(\cos\theta_{\rm j}-\cos\theta_{\rm j+1}\right). 
\end{eqnarray}

%%%%%%%%%%%%%%%%%%%%%%%%%%%%%%%%%%%%%%%%%%%%%%%%%%

% Don't change these lines
\bsp	% typesetting comment
\label{lastpage}
\end{document}